\documentclass[10pt,twocolumn,preprintnumbers,amsmath,amssymb,nofootinbib
,superscriptaddress]{revtex4-2}
\usepackage{graphicx,longtable,mathrsfs,color,array}
\usepackage[colorlinks,plainpages]{hyperref}
\hypersetup{urlcolor=blue,citecolor=blue, linkcolor=blue,hypertexnames=false}
\usepackage[usenames,dvipsnames]{xcolor} 
\usepackage{amssymb,amsmath,mathtools,mathrsfs,slashed} 
\usepackage{epsfig,subfigure,placeins,float} 
\usepackage{booktabs,longtable,ctable,multirow} 
\usepackage{exscale,relsize} 
\usepackage[normalem]{ulem} 
\usepackage{enumerate}
\usepackage{enumitem}
\usepackage{comment}
\usepackage{color}
\allowdisplaybreaks[1]

\newcommand{\be}{\begin{equation}}
\newcommand{\ee}{\end{equation}}
\newcommand{\bea}{\begin{eqnarray}}
\newcommand{\eea}{\end{eqnarray}}

\newcommand{\gr}{{\mbox{\tiny GR}}}

\newcommand{\pv }{{\mbox{\tiny PV}}}
\newcommand{\cs }{{\mbox{\tiny CS}}}
\newcommand{\rl }{{\mbox{\tiny R,L}}}
\newcommand{\ri }{{\mbox{\tiny R}}}
\newcommand{\lef }{{\mbox{\tiny L}}}
\newcommand{\uni }{{\mbox{\tiny uniform}}}
\newcommand{\delayed }{{\mbox{\tiny delayed}}}
\newcommand{\sfr}{{\mbox{\tiny SFR}}}

\newcommand{\UniformRateKappaDcMarginal}{-0.01^{+0.25}_{-0.28}}
\newcommand{\UniformRateKappaZMarginal}{0.03^{+0.54}_{-0.55}}
\newcommand{\UniformRateKappaDcConditioned}{0.01^{+0.10}_{-0.12}}
\newcommand{\UniformRateKappaZConditioned}{0.01^{+0.21}_{-0.22}}

\newcommand{\SfrKappaDcMarginal}{0.01^{+0.16}_{-0.18}}
\newcommand{\SfrKappaZMarginal}{-0.01^{+0.28}_{-0.28}}
\newcommand{\SfrKappaDcConditioned}{0.01^{+0.09}_{-0.11}}
\newcommand{\SfrKappaZConditioned}{-0.01^{+0.18}_{-0.17}}

\newcommand{\DelayedSfrKappaDcMarginal}{0.01^{+0.16}_{-0.18}}
\newcommand{\DelayedSfrKappaZMarginal}{-0.02^{+0.26}_{-0.25}}
\newcommand{\DelayedSfrKappaDcConditioned}{0.01^{+0.09}_{-0.10}}
\newcommand{\DelayedSfrKappaZConditioned}{-0.01^{+0.15}_{-0.15}}

\newcommand{\VariableRateKappaDcMarginal}{0.01^{+0.16}_{-0.17}}
\newcommand{\VariableRateKappaZMarginal}{-0.01^{+0.27}_{-0.27}}
\newcommand{\VariableRateKappaDcConditioned}{-0.00^{+0.11}_{-0.11}}
\newcommand{\VariableRateKappaZConditioned}{-0.00^{+0.20}_{-0.18}}

\newcommand{\MaxLikelihoodKappaDC}{0.107}
\newcommand{\MaxLikelihoodKappaZ}{-0.010}
\newcommand{\HLSNR}{1.2}
\newcommand{\HVSNR}{2.0}
\newcommand{\LVSNR}{-0.7}

\newcommand{\ProbHVSNR}{0.044}

\newcommand{\ProbHVLVSNR}{0.0065}
\newcommand{\ProbAboveWithTrials}{0.20}
\newcommand{\ProbAboveAndBelowWithTrials}{0.10}
\newcommand{\ProbAboveWithTrialsNoise}{0.11}
\newcommand{\ProbAboveAndBelowWithTrialsNoise}{0.07}

\begin{document}

\title{A New Probe of Gravitational Parity Violation \\ Through (Non-)Observation of the Stochastic Gravitational Wave Background}

\author{Thomas Callister}
\email{tcallister@uchicago.edu}
\affiliation{Kavli Institute for Cosmological Physics, University of Chicago,  Chicago, IL 60637, USA}
\author{Leah Jenks}
\email{ljenks@uchicago.edu}
\affiliation{Kavli Institute for Cosmological Physics, University of Chicago,  Chicago, IL 60637, USA}
\author{Daniel E. Holz}
\affiliation{Kavli Institute for Cosmological Physics, University of Chicago,  Chicago, IL 60637, USA}
\affiliation{Enrico Fermi Institute, University of Chicago,  Chicago, IL 60637, USA}
\affiliation{Department of Physics, University of Chicago,  Chicago, IL 60637, USA}
\affiliation{Department of Astronomy and Astrophysics, University of Chicago,  Chicago, IL 60637, USA}
\author{Nicol\'as Yunes}
\email{nyunes@illinois.edu}
\affiliation{Illinois Center for Advanced Studies of the Universe, Department of Physics, University of Illinois at Urbana-Champaign, Urbana, Illinois 61801, USA.}

\date{Received \today; published -- 00, 0000}

\begin{abstract}
Parity violation in the gravitational sector is a prediction of many theories beyond general relativity. In the propagation of gravitational waves, parity violation manifests by inducing amplitude and/or velocity birefringence between right- and left-circularly polarized modes. We study how the stochastic gravitational wave background can be used to place constraints on these birefringent effects. We consider two model scenarios, one in which we allow birefringent corrections to become arbitrarily large, and a second in which we impose stringent theory priors. In the former, we place constraints on a generic birefringent gravitational-wave signal due to the current non-detection of a stochastic background  from compact binary events. We find a joint constraint on birefringent parameters, $\kappa_D$ and $\kappa_z$, of $\mathcal{O}(10^{-1})$. In the latter scenario, 
we forecast constraints on parity violating theories resulting from observations of the future upgraded LIGO-Virgo-KAGRA network as well as proposed third-generation detectors. We find that third-generation detectors will be able to improve the constraints by at least two orders of magnitude, yielding new stringent 
bounds on parity violating theories. 
This work introduces a novel and powerful probe of gravitational parity violation with gravitational-wave data.

\end{abstract}

\date{\today}
\maketitle


\section{Introduction}
The era of gravitational-wave astrophysics has thus far yielded $\mathcal{O}(100)$ observations of compact binary coalescences (CBCs) from the LIGO-Virgo-KAGRA (LVK) collaboration~\cite{TheLIGOScientific:2014jea,TheVirgo:2014hva,KAGRA:2020tym,LIGOScientific:2021djp}. Moreover, pulsar timing arrays have provided significant evidence for a low-frequency stochastic gravitational wave background, thought to arise from merging supermassive black holes~\cite{NANOGrav:2023gor,NANOGrav:2023hvm,EPTA:2023fyk,Xu:2023wog,Reardon:2023gzh}.
The stellar-mass compact binaries observed by the LVK collaboration are additionally expected to produce a high-frequency stochastic background, due to the cumulative signal of sources too distant to be individually resolved.
As of the third 
LVK observing run (O3), this stochastic background has not been detected, and current sensitivity projections indicate that a detection in the ongoing fourth observing run (O4) is unlikely~\cite{KAGRA:2021kbb,KAGRA:2021mth}.  

Gravitational wave observations serve as a test for a variety of questions regarding fundamental physics and the nature of gravity~\cite{LIGOScientific:2020tif,LIGOScientific:2021sio}.
One such question is the existence of gravitational parity violation:
Do gravitational interactions remain invariant under a reversal of spatial coordinates?
Parity is a fundamental symmetry, but we know from observation that our universe is not parity symmetric; for example, the weak force is known from experiments to be chiral~\cite{Lee:1956qn,Wu:1957my}.
Furthermore, parity violation has been recently hinted at in observations of galaxy four-point functions \cite{Hou:2022wfj, Philcox:2022hkh}.
General Relativity (GR) is a parity-even theory, but parity violation can arise in theories that are extensions to GR.
Chern-Simons (CS) gravity is a notable example \cite{jackiw, Alexander:2009tp}, with others including variants of CS gravity, parity-violating extensions of ghost-free scalar-tensor theory, symmetric teleparallel gravity, and Horava-Lifshitz theory \cite{Sulantay:2022sag, Bombacigno:2022naf, Boudet:2022nub, Nojiri:2019nar, Crisostomi:2017ugk, Conroy:2019ibo, Horava:2009uw,Zhu:2013fja}. In general, parity-violating extensions of GR arise as low-energy effective field theories of some higher-energy, UV theory (e.g. \cite{Polchinski:1998rr, ALVAREZGAUME1984269, PhysRevLett.96.081301, Green:1987mn, Alexander:2004xd, Alexander:2021ssr, Alexander:2022cow}).

Gravitational parity violation can lead to modifications to both the generation of gravitational waves in compact binaries and to their propagation~\cite{Alexander:2007kv,Yagi:2013mbt}.
In this context, parity-violating modifications to gravitational-wave propagation lead to \textit{birefringence}, in which the right- and left-circular polarization modes evolve differently in their phase and/or amplitude~\cite{Lue:1998mq, Alexander:2004wk, Alexander:2007kv, Yunes:2010yf,Alexander:2017jmt,Ezquiaga:2021ler, Jenks:2023pmk}. 
The amplitude modifications, which we will refer to as ``amplitude birefringence,'' force one of the polarizations to grow with propagation distance, while the other decays. 
The phase modifications, which we will refer to as ``velocity birefringence,'' force a frequency-dependent phase-shift that grows with distance and has different signs depending on the polarization state. 
Such birefringence effects leave a signature that gravitational-wave detectors can, in principle, observe, as first suggested in~\cite{Alexander:2007kv,Yunes:2010yf}. In what follows we specifically focus on amplitude birefringence.

Gravitational-wave birefringence due to parity violation has been searched for in a variety of ways in the context of ground-based detectors.
The $\mbox{GWTC-2}$ and GWTC-3 catalogs have been used to place constraints on amplitude birefringence by stacking individual constraints from binary black holes (BBHs)~\cite{Okounkova:2021xjv,Ng:2023jjt}.
Both amplitude and velocity birefringence  have been constrained using coincident gravitational waves and gamma rays from the binary neutron star (BNS) event GW170817~\cite{BNSPV}.
Velocity birefringence has also been constrained using the GWTC-3 catalog \cite{Wang:2020cub, Zhao:2022pun}. Several studies have made projections for the possibility of probing parity violation from inflationary physics in the stochastic background with ground-based and future detectors \cite{Seto:2007tn,Seto:2008sr,Cai:2021uup, Orlando:2020oko}, and have placed upper limits on primordial parity violation with LVK observations \cite{Crowder:2012ik, Martinovic:2021hzy,Jiang:2022uxp}. Additionally, projections have been made for the prospect of using future observations of the stochastic background to probe parity violation from compact binaries \cite{Yagi:2017zhb}, but as of yet, no constraint of this type exists in the literature with current data.

\begin{figure}[t]
    \centering
 \includegraphics[width=0.48\textwidth]{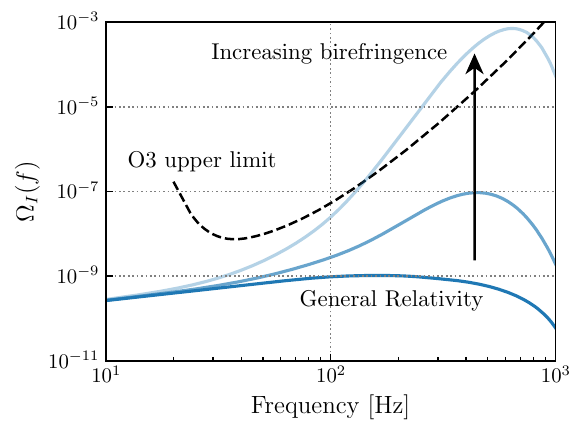}
    \caption{
    Illustrative scenario of our test of gravitational-wave birefringence.
    The lower-most blue curve shows a prediction for the total energy-density (i.e. the Stokes-$I$ energy density) of the stochastic background arising from binary black hole mergers in General Relativity.
    Increasing birefringence enhances the amplitude of the stochastic background to a point at which it is excluded by current non-observation~\cite{KAGRA:2021kbb}.
    }
    \label{fig:birefringence_simple_example}
\end{figure}

The above work all seeks to test gravitational-wave birefringence via the detection of compact binary mergers.
Our objective for this work is different: we seek to instead use the \textit{non-detection} of a stochastic gravitational-wave background to constrain amplitude birefringence.
The idea is simple:
By virtue of amplifying one polarization, sufficiently strong amplitude birefringence will increase the net energy in the gravitational-wave background (even as it suppresses the other polarization).
If there exists a modification that enhances the gravitational-wave amplitude of distant compact binaries, then the size of that modification must therefore be limited by the fact that we have observed no stochastic background from these objects.
The stochastic background is a particularly powerful constraining tool for amplitude birefringence because the degree of gravitational-wave amplification grows with propagation distance.
The stochastic background comprises the cumulative signal of merging binaries out to a redshift of $z \approx 10$, which because of their distance experience birefringent enhancements far greater than any single directly-observable event in the local Universe.
Thus, even without an observation of the gravitational-wave background, we can place a competitive upper bound on the magnitude of any parity violating modification to GR. This analysis, therefore, is distinct from that of individual compact binary mergers and provides an independent test of parity violation. 
We note that, although we focus here on the (undetected) background arising from compact binaries, the same logic can be equivalently applied to undetected backgrounds arising from any class of gravitational wave source

Figure~\ref{fig:birefringence_simple_example} describes the idea through a toy example.
Shown in blue are several predictions for the amplitude, $\Omega_I(f)$, of the stochastic gravitational-wave background (i.e.~the dimensionless energy density associated with the Stokes-$I$ parameter, the total strain power across both gravitational-wave polarizations; see Eq.~\eqref{eq:OmgI}) from stellar-mass binary black holes as a function of gravitational-wave frequency, $f$.
For this toy example, we adopt a fixed black hole mass function and merger history, to be described later below.
The bottom curve, labeled ``General Relativity,'' shows the anticipated gravitational-wave background according to GR, in the absence of amplitude birefringence.
This predicted signal lies at least an order of magnitude below present-day search sensitivities~\cite{KAGRA:2021kbb}, illustrated with the dashed black curve, consistent with the lack of detection during O3.
The existence of parity-violating amplitude birefringence, though, would boost the expected gravitational-wave background, as we shall show in this paper in detail.
Any degree of birefringence that were  to boost the expected background above present-day limits, as in the hypothetical case of the uppermost blue curve, would then be excluded.

This paper will explore constraints on parity-violating amplitude birefringence under the following 
complementary approaches:
    \begin{enumerate}
    \item \textit{Phenomenological approach}:
    We here proceed 
    agnostically by not imposing strongly-informative theory-motivated priors on the size of possible parity-violating effects. 
    Instead, we adopt entirely uninformative priors that allow gravitational-wave amplification to become large.
    \item \textit{Theory-motivated approach}:
    We here impose strong, theoretically-motivated priors. 
    Under this approach, allowable amplifications must be sufficiently small to remain well-described as linear deformations of GR.
    \end{enumerate}
In the phenomenological approach, right-polarized gravitational waves are exponentially enhanced, while left-polarized ones are exponentially suppressed, leading to a net enhancement of the stochastic background. In the theory-motivated approach, however, the exponential enhancement and suppression must be Taylor-expanded, and thus, the amplitude of the stochastic background  is not modified to linear order. In this case, sensitivity to birefringence arises from the polarization asymmetry of the background, rather than an overall enhancement.

In the phenomenological approach, we calculate the enhancement of the stochastic background  signal for three fixed models of the BBH merger rate and find a joint constraint on the two parameters that control amplitude birefringence, $\kappa_D$ and $\kappa_z$. Given that different models of the merger rate can alter the ampltiude of the stochastic background, we then drop this assumption, and instead, we simultaneously infer $\kappa_D$ and $\kappa_z$ along with the best-fit merger rate. Through this analysis, we find that both $\kappa_D$ and $\kappa_z$ must be smaller than $\lesssim \mathcal{O}(10^{-1})$. We then turn to the theory-motivated approach and impose theory-motivated priors on the allowable size of the birefringent modification. In this case, the maximum stochastic background enhancement is well below the O3 sensitivity. Therefore, we forecast how future gravitational-wave detectors with increased sensitivity may be able to place competitive constraints in this regime. We find that third-generation detectors, such as Cosmic Explorer or Einstein Telescope, are likely to reach the sensitivity needed to place constraints of $\mathcal{O}(10^{-3})$ on the birefringent parameters of parity-violating theories.

The remainder of this paper presents the details that lead us to the results summarized above.
In Sec.~\ref{sec:parity-violation} we discuss relevant details of general parity-violating gravity theories and how gravitational-wave polarizations are birefringently altered. 
In Sec.~\ref{sec:SGWB} we review the stochastic gravitational-wave background from binary black holes and show, under each of our above approaches, how it is modified in parity violating gravity.
In Sec.~\ref{sec:methods} we describe our inference methods, and in Secs.~\ref{sec:phenomenological-results} and~\ref{sec:theory-motivated-results} we present results under both the phenomenological and theory-motivated approaches, respectively.
We conclude in Sec.~\ref{sec:discussion}. Unless explicitly noted otherwise, in what follows we work in geometric units such that $G=1=c$, and assume a plus metric signature (-, +, +, +). 

\section{Gravitational Parity Violation}
\label{sec:parity-violation}

Numerous theories that go beyond GR and lead to parity violation have been proposed.
Theories that contain parity violation are typically motivated from a high-energy theory, which leads to small deviations from GR at low energies. 
In general, one can characterize a theory that includes parity violation via the action
\be 
S = S_\gr + S_\pv(R, \vartheta) , 
\ee 
where $S_\gr$ is the usual Einstein-Hilbert action of GR,
\be 
S_\gr  = \int d^4x \sqrt{-g} R,
\ee 
with $R$ the Ricci tensor and $g$ the metric determinant.
The quantity $S_\pv(R,\vartheta)$ is the parity violating contribution, which is, in general, a function of parity-violating curvature invariants and an auxiliary pseudo-scalar field, denoted by $\vartheta$.
The most well studied parity violating theory is Chern-Simons gravity \cite{jackiw, Alexander:2009tp}, in which an auxiliary pseudo-scalar field couples to the (parity odd) Pontryagin density of spacetime. 
However, parity-violating interactions can also be constructed in other beyond-GR theories, such as in ghost-free scalar-tensor gravity, the symmetric teleparallel equivalent of GR, and Horava-Lifshitz gravity \cite{Sulantay:2022sag, Bombacigno:2022naf, Boudet:2022nub, Nojiri:2019nar, Crisostomi:2017ugk, Conroy:2019ibo, Horava:2009uw,Zhu:2013fja}.

As a representative example, consider the Chern-Simons action, where 
\be 
S_\pv^{\cs} = \int d^x \sqrt{-g} \left[\frac{\alpha}{4\kappa}\vartheta \; *RR- \frac{1}{2} \left(\partial_{\mu} \vartheta\right) \left(\partial^{\mu} \vartheta\right)\right], 
\label{eq:S-CS}
\ee 
where $\kappa = (16\pi)^{-1}$, $\alpha$ is the Chern-Simons coupling parameter, and $*RR$ is the Pontryagin density of the spacetime, defined by 
\be 
*RR = \frac{1}{2}\epsilon^{cdef}R^a{}_{bef}R^b{}_{acd}, 
\ee 
with $\epsilon^{cdef}$ the totally antisymmetric Levi-Civita tensor such that $*R$ is the Hodge dual of the Riemann tensor. The pseudo-scalar field, $\vartheta$, satisfies the wave equation
\be 
\Box \vartheta = -\frac{\alpha}{4\kappa} *RR\,, 
\label{eq:boxtheta}
\ee 
and this pseudo-scalar also enters the modified field equations, which can be obtained by varying the above action with respect to the metric. 
For other parity violating theories, the specific forms of the action differ, by e.g.~including higher derivative curvature invariants and more complicated functions of $\vartheta$, but the same fundamental principles apply. In this work, we will \textit{not} assume a specific form for $S_\pv$ and attempt to remain mostly theory-agnostic throughout the remainder of our analysis.

Both the generation and propagation of gravitational waves are impacted in parity-violating theories~\cite{Alexander:2007kv,Yagi:2013mbt}.
Although each of these effects is related to the presence of an auxiliary pseudo-scalar field, they each arise from \textit{markedly} different aspects of the theory. 
For maximum clarity we describe the effects in the context of CS gravity as a representative example, but note that the discussion below holds generically for parity violating gravity theories. 

First consider gravitational-wave generation.
Parity violating generation effects arise from the fact that the psuedo-scalar field extracts energy and momentum from the binary, changing the (dissipative) radiation-reaction force. Generation effects also arise because the spacetime of black holes and neutron stars is modified due to the pseudo-scalar field, changing the Hamiltonian, and thus, the (conservative) binding energy of binary systems. These modifications combine to introduce corrections to the binary inspiral and chirping rate, which, in turn, lead to corrections to both the amplitude and phase of the gravitational wave, as described in \cite{Alexander:2017jmt}.
In the case of CS gravity, for example, the result is a 2PN correction to both the phase \cite{Yagi:2012vf} and the amplitude \cite{Tahura:2019dgr} of the gravitational waves emitted. 
One key point to emphasize is that the above modifications to gravitational-wave generation arise from the \textit{particular} solution of Eq.~\eqref{eq:boxtheta}, $\vartheta_P$, namely a pseudo-scalar field that is sourced by the spacetime curvature itself due to a non-zero Pontryagin density (see e.g.~\cite{Yunes:2009hc}). In this case, the pseudo-scalar $\vartheta_P$ is proportional to the coupling constant, $\alpha$, and therefore, since the parity-violating action is also proportional to $\alpha$ (see e.g.~Eq.~\eqref{eq:S-CS}), the corrections to the amplitude and phase of gravitational waves are $\mathcal{O}(\alpha^2)$. 

Let us now consider the propagation of gravitational waves. In this case, parity violation can lead to both real and imaginary corrections to the gravitational-wave phase.
To see this more clearly, imagine a circularly polarized signal in the right and left basis.
Imaginary corrections to the phase lead to an overall birefringence in the amplitude of the wave, amplifying one of the polarization modes and attenuating the other.
Similarly, the real corrections to the phase lead to birefringent contributions to the true phase and dispersion relation of the gravitational waves.
This leads to velocity birefringence, in which the right- and left-circular modes propagate at different speeds.
In contrast to the generation effects, propagation effects arise due to the \textit{homogeneous} solution of Eq.~\eqref{eq:boxtheta}, $\vartheta_H$. In this case, we are no longer considering the particular scalar field that is sourced by spacetime curvature, $\vartheta_P$, but rather a generic background field $\vartheta_H$ that shares the symmetries of the cosmological Friedmann-Robertson-Walker (FRW) background, as described in~\cite{Alexander:2017jmt}. In particular, the background pseudoscalar $\vartheta_H$ inducing propagation effects is \text{not} the same as the Pontryagin-sourced scalar that induces generation effects $\vartheta_P$. Furthermore, in this case, the modifications are proportional to $\alpha\dot{\vartheta}_H$, in contrast to the parity violating generation modifications, which enter at $\mathcal{O}(\alpha^2)$ and are not dependent on the $\dot{\vartheta}_H$ of the background field.

Thus, we can generically consider the effects on generation and propagation separately, as they arise from distinct dynamics. Furthermore, any constraint on $\alpha$ due to generation modifications does not directly constrain propagation modifications, because the propagation also depends on $\dot{\vartheta}_H$. For example, $\alpha$ could be tightly constrained from effects that arise in the generation of gravitational waves, but $\dot{\vartheta}_H$ need not be similarly small in the propagation sector. The generation effects also tend to be subdominant, due to their entry at $\mathcal{O}(\alpha^2)$, while propagation effects are ${\cal{O}}(\alpha)$ and grow with distance traveled.    

The effects of parity violation on gravitational-wave propagation are most easily described in the circular polarization basis.
Let $\tilde{h}^\gr_\rl(f)$ be the Fourier transform of the gravitational-wave strain in right- and left-circular polarization modes, as normally expected in GR. $\tilde{h}^\gr_\rl$ are related to the linearly polarized modes via 
\begin{align}
    \tilde{h}_+ &= \frac{\tilde{h}_R + \tilde{h}_L}{\sqrt{2}}, \qquad
    \tilde{h}_\times = i\frac{\tilde{h}_R - \tilde{h}_L}{\sqrt{2}}. 
\end{align}
The effects of parity violation during gravitational-wave propagation can be most generally characterized in the following theory-agnostic way \cite{Jenks:2023pmk}:
\begin{align}
&\tilde{h}_\rl(f) = \tilde{h}^\gr_\rl(f)\nonumber \\
&\times \exp\left\{\mp\sum_n\frac{\left[k(1+z)\right]^n}{2}\left(\frac{\alpha_{n}}{\Lambda_\pv ^n}z_n + \frac{\beta_{n}}{\Lambda_\pv ^{n-1}} D_{n+1}\right)\right\}\nonumber \\
&\times \exp\left\{\pm i\sum_m\frac{\left[k(1+z)\right]^m}{2}\left( \frac{\gamma_{m}}{\Lambda_\pv ^m}z_m + \frac{\delta_{m}}{\Lambda_\pv ^{m-1}}D_{m+1}\right)\right\},
\label{hRLmod}
\end{align}
where $k=2\pi f$ is the wavenumber, $\alpha, \beta,\gamma,$ and $\delta$ parameterize the parity violation\footnote{In \cite{Jenks:2023pmk}, these coefficients are denoted $\alpha_{n_0}$, etc., because we assume that they are well described by their present-day values. Throughout this work, we make the same assumption, but drop the $0$ subscript for clarity.}, $n$ is an odd integer, and $m$ is an even integer.
The first (real) exponential term in Eq.~\eqref{hRLmod} corresponds to corrections to the amplitude, while the second (imaginary) term corresponds to corrections to the phase.
We take $\Lambda_\pv$ to be the cutoff scale of the effective field theory that introduces parity-violating effects, and also define the effective distance, $D_\alpha$, and effective redshift, $z_\alpha$, such that
\be 
D_\alpha = (1 +z)^{1-\alpha}\int \frac{(1 + z)^{\alpha-2}}{H(z)}dz,
\ee 
and 
\be 
z_\alpha = (1+z)^{-\alpha}\int \frac{dz}{(1 + z)^{1 - \alpha}}.
\ee 
Here, $H(z)$ is the Hubble parameter as a function of redshift.
Notice that $z_0 = \ln(1+z)$, $z_1 = z/(1+z)$, $D_1 = D_T$, and $D_2 = D_A$, where $D_T$ is the look-back or light-travel distance and $D_A$ is the angular diameter distance, all of which coincides at small redshift, i.e., $z_0 \sim z_1 \sim z_\alpha \sim z$ and $D_1 \sim D_2 \sim D_A$ for $z \ll 1$, but not at large redshift~\cite{Jenks:2023pmk}. 
The parametrization presented above was derived from generic symmetry principles, and found to map explicitly to the predictions of a large number of parity-violating modified gravity theories~\cite{Jenks:2023pmk}. 

Let us now simplify the above generic parameterization by considering a subset of cases.
While there is in principle an infinite sum over odd $n$ and even $m$, in practice modified gravity theories only predict that a small subset is non-zero. For example, Chern-Simons gravity corresponds to $\alpha_{1} \neq 0$ with all other coefficients vanishing. Given this, we will be interested in parity violating corrections of the general form
\be 
\tilde{h}_{\rl}(f) = \tilde{h}^\gr_\rl(f) e^{\mp v(f)}, 
\label{eq:hvw}
\ee 
where $v(f)$ is the parity-violating function of frequency drawn from the arguments of the real exponential in Eq.~\eqref{hRLmod}. 

\subsubsection*{Approach 1: Phenomenological}

In the phenomenological approach, we take inspiration from Eq.~\eqref{hRLmod} but do not use it exactly, because as mentioned, we do not consider theoretical restrictions. Specifically, we consider circularly-polarized gravitational waves in the form of Eq.~\eqref{eq:hvw}, where $v(f)$ characterizes the amplitude birefrigence contributions. We consider only the lowest-order parity-violating effects, in analogy to $n=1$ in Eq.~\eqref{hRLmod}. We thus define $v_{\text{p}}(f)$ to be 
\be 
v_{\text{p}}(f) = \frac{\pi f}{100 \text{Hz}}\left(\kappa_z z + \kappa_D\frac{D_C}{\text{Gpc}}\right),
\label{eq:birefringent-v}
\ee 
where $D_C$ is the comoving distance, and for convenience we have scaled $f$ and $D_C$ by benchmark values.
 We have introduced the (dimensionless) parity violating parameters $\kappa_z$ and $\kappa_D$, to denote the corrections that scale with redshift and distance, respectively. These correspond to $\alpha_{1}$ and $\beta_{1}$ in Eq.~\eqref{hRLmod}, respectively.

\subsubsection*{Approach 2: Theory Motivated}

We now examine a modification grounded in beyond-GR parity violating theories, and consider the leading $n=1$ contributions from Eq.~\eqref{hRLmod} for the amplitude modification. In this case, we define $v_{\text{th}}(f)$ as 
\be 
v_{\text{th}}(f) = \pi f \left(\frac{\alpha_{1}}{\Lambda_{\pv}}z  + \beta_{1} D_C\right), 
\label{eq:eft-v}
\ee 
where $\alpha_{1}$ and $\beta_{1}$ are the theory parameters appearing in Eq.~\eqref{hRLmod} and $\Lambda_\pv$ is the parity violating cutoff scale. Recall that in geometric units $f \Lambda_\pv^{-1}$ and $f D_C$ are unitless quantities such that $\alpha_1$ and $\beta_1$ are also unitless.
In this case, we assume that any deviations from GR must be small, and thus expand the waveform in small $v(f)$ such that 
\be
h_\rl \approx h^\gr_\rl\left[ 1 \mp v_{\text{th}}(f) + \mathcal{O}(v^2)\right]. 
\label{eq:hrlLinear}
\ee 
The prior $v(f) \ll 1$ changes as the gravitational wave evolves in frequency and distance. To remain consistent with a small-deformation expansion, we will take the most conservative bound, such that $v(f_{\max}, z_{\max}) \ll 1$. We discuss this in further detail in Sec.~\ref{sec:theorypriors}.
 
Once a constraint on $v$ is obtained, one can then easily map it to constraints on coupling constants of a variety of parity-violating theories, following~\cite{Jenks:2023pmk}. 
In fact, Eq.~\eqref{eq:hrlLinear} can be mapped cleanly to a parameterized post-Einsteinian (ppE) waveform \cite{PPE,PhysRevD.57.2061, Mirshekari:2011yq, Chatziioannou:2012rf, Sampson:2013wia, cornishsampson}, as explicitly shown in \cite{Jenks:2023pmk}.

\section{The Stochastic Gravitational Wave Background}
\label{sec:SGWB}

For every detected compact binary merger, there are many more mergers occurring in the distant Universe, beyond the horizon of present-day detectors.
Although individually unresolvable, the superposition of all such sources gives rise to an astrophysical stochastic gravitational-wave background~\cite{Regimbau:2011rp,Christensen:2018iqi,KAGRA:2021kbb}.
This background is, in principle, detectable in the form of excess \textit{correlations} between pairs of gravitational-wave detectors.
In Sec.~\ref{sec:sgwb-intro} we describe the stochastic background  as predicted by general relativity.
In Sec.~\ref{sec:nongr-sgwb-intro} we describe how this picture is altered in the presence of parity-violating amplitude birefringence. Then, in Sec.~\ref{sec:illustration}, we qualitatively illustrate the effects of the birefringent corrections to the gravitational-wave background. 

\subsection{The Stochastic Gravitational-Wave Background in General Relativity}
\label{sec:sgwb-intro}

Consider a plane wave expansion of the gravitational-wave content at a given location, where $\tilde h_\rl (f,\hat n)$ is the Fourier-transformed, right- or left-circular polarization modes of gravitational waves with (detector-frame) frequency $f$, propagating in direction $\hat n$.
~Assuming that the astrophysical gravitational-wave background is stationary, Gaussian, and isotropic, the covariance between the strains $\tilde h_\rl (f,\hat n)$ and $\tilde h_\rl (f',\hat n')$ measured at two different frequencies and from two directions is~\cite{Allen:1997ad}
    \be 
    \langle \tilde h_\rl (f, \hat n)\tilde h^*_\rl (f',\hat n')\rangle = \frac{\delta(f - f')}{2}\frac{\delta^2(\hat n,\hat n')}{4\pi} \mathcal{H}_\rl (f)\,,
    \label{eq:strain-power}
    \ee 
where $\left<\cdot\right>$ stands for a time average over a duration much longer than individual gravitational-wave signals (or, equivalently, an ensemble average over many realizations of the gravitational-wave sky assuming ergodicity). 
This equation can be taken as the \textit{definition} of $\mathcal{H}_{\rl}(f)$, the one-sided strain power-spectral density of the gravitational-wave background in right- or left-circular polarization modes.
In practice, it is more common to describe the stochastic background  instead by the dimensionless energy densities in each polarization, related to their strain power-spectral densities via~\cite{Allen:1997ad}
\be
\Omega_\rl(f) = \frac{2\pi^2}{3 H_0^2} f^3\,\mathcal{H}_\rl(f),
\label{eq:energy-right-left}
\ee
where $H_0$ is the Hubble constant.

To compute the total energy density of the stochastic gravitational-wave background, we integrate over the contributions from individual sources at all redshifts.
Let $\langle dE_{\rl}/df\rangle_s$ be the source-frame energy spectrum radiated by a single compact binary in right- or left-circular polarizations, averaged across the compact binary population.
If $R_m(z)$ is the comoving merger rate density of these sources, then the total energy densities in right- and left-circular modes can be shown to be~\cite{Allen:1997ad,Phinney:2001di} 
\be 
\begin{aligned}
{\Omega}_{\rl} (f) = \frac{f}{H_0\rho_c}\int dz &\frac{R_m(z)}{(1 + z)\sqrt{\Omega_m(1 + z)^3 + \Omega_\Lambda}} \\
    &\times \left\langle \frac{dE_\rl}{df} \right\rangle_s \Big|_{f(1+z)},
\end{aligned}
\label{eq:RL-gr}
\ee
where recall that $\langle \rangle_s$ denotes an average over all binary sources at the given redshift, and $\Omega_{m,\Lambda}$ are the energy densities of matter and dark energy, respectively, as compared to the critical density of the universe, $\rho_c$. In this paper, we will use the highest-likelihood, Planck 2018 values $\Omega_m = 0.31$, $\Omega_\Lambda = 0.69$ and $\rho_c = 8.6 \times 10^{-30}$ g/cm${}^3$ \cite{Planck:2018vyg}, but these choices do not affect the conclusions we will obtain qualitatively.   
Moreover, even though the posterior probability distributions obtained by Planck are not delta functions, statistical errors are small enough that they will not affect any of the conclusions of this paper.   
Note that the radiated energy-spectrum is evaluated not at the detector-frame $f$, but at the appropriately \textit{blue-shifted} source-frame frequency $f_{\rm src} = f (1+z)$. 

In addition to this right- and left-circular polarization basis, we can alternatively (and equivalently) consider a stochastic background in terms of the Stokes $I$ and $V$ parameters.
In this Stokes parameter picture, we rewrite the plane wave expansion as~\cite{Seto:2007tn,Seto:2008sr,Crowder:2012ik,Jiang:2022uxp}
\be 
\langle \tilde h_\rl (f, \hat n)\tilde h^*_\rl (f',\hat n')\rangle = \frac{\delta(f - f')}{2}\frac{\delta^2(\hat n,\hat n')}{4\pi}[I(f) \pm V(f)].
\ee 
The Stokes parameter $I(f) = \mathcal{H}_\ri(f) + \mathcal{H}_\lef(f)$ gives the total strain power across both polarizations.
The Stokes parameter $V(f) = \mathcal{H}_\ri(f) - \mathcal{H}_\lef(f)$ quantifies the difference between the two polarizations, with positive or negative $V(f)$ indicating an excess of right- or left-circular modes, respectively.
Proceeding analogously to Eq.~\eqref{eq:energy-right-left}, we can define dimensionless energy densities
\be
\Omega_I(f) = \frac{2\pi^2}{3 H_0^2} f^3 I(f)
\ee
and
\be
\Omega_V(f) = \frac{2\pi^2}{3 H_0^2} f^3 V(f)
\ee
associated with each Stokes parameter.
These Stokes energy densities are related to the circular polarization energy densities in Eq.~\eqref{eq:energy-right-left}
by
    \begin{equation}
    \Omega_I(f) = \Omega_R(f) + \Omega_L(f)
    \label{eq:OmgI}
    \end{equation}
and
    \begin{equation}
    \Omega_V(f) = \Omega_R(f) - \Omega_L(f).
    \label{eq:OmgV}
    \end{equation}
The quantity $\Omega_I(f)$ represents the total energy of the gravitational-wave background, while $\Omega_V(f)$ is the difference between energy densities in right- and left-circular polarizations.
Accordingly, $\Omega_I(f)$ is strictly positive but $\Omega_V(f)$ can be either positive or negative, depending on whehter right- or left-circular polarizations dominate.
Using Eq.~\eqref{eq:RL-gr}, the energy densities in each Stokes parameter are given by
\be 
\begin{aligned}
{\Omega}_{I} (f) = \frac{f}{H_0\rho_c}\int dz &\frac{R_m(z)}{(1 + z)\sqrt{\Omega_m(1 + z)^3 + \Omega_\Lambda}} \\
    &\times \left\langle\frac{dE_\ri}{df}+\frac{dE_\lef}{df}\right\rangle_s \Big|_{f(1+z)}.
\end{aligned}
\label{eq:stokes-I-gr}
\ee
and
\be 
\begin{aligned}
{\Omega}_{V} (f) = \frac{f}{H_0\rho_c}\int dz &\frac{R_m(z)}{(1 + z)\sqrt{\Omega_m(1 + z)^3 + \Omega_\Lambda}} \\
    &\times \left\langle\frac{dE_\ri}{df}-\frac{dE_\lef}{df}\right\rangle_s \Big|_{f(1+z)}.
\end{aligned}
\label{eq:stokes-V-gr}
\ee
In the absence of birefringence (and if gravitational-wave sources are isotropically oriented), then on average we expect equal amounts of radiation in right- and left-circular modes, such that $\Omega_\ri(f) = \Omega_\lef(f) = \frac{1}{2}\Omega_I(f)$.
Accordingly, the Stokes $V$ energy density, $\Omega_V$, should be identically zero on average.

\subsection{Parity Violating Modifications}
\label{sec:nongr-sgwb-intro}

The presence of parity violating gravitational-wave birefringence would alter the above picture, however, breaking the symmetry between right- and left-circular polarizations and amplifying both $\Omega_I(f)$ and $\Omega_V(f)$.
As we have discussed previously, we will take two different approaches towards modifying the gravitational-wave background in the presence of amplitude birefringence: a \textit{phenomenological} and a \textit{theory-motivated} approach. Below, we enumerate how the stochastic background is altered in each approach. 

\subsubsection*{Approach 1: Phenomenological}
\label{sec:phenom-amplification}

We first take a purely phenomenological approach, modifying $\Omega_I(f)$ and $\Omega_V(f)$ to reflect possible (arbitrarily large) exponential amplification of right- or left-circular modes. The waveform modification in this approach is given by Eq.~\eqref{eq:hvw}, where $v(f)$ is given by Eq.~\eqref{eq:birefringent-v}. Then, the ensemble-averaged radiated energy per unit gravitational-wave frequency is 
\begin{equation}
\begin{aligned}
    \left\langle\frac{dE_\ri}{df}+\frac{dE_\lef}{df}\right\rangle_s
        &\propto f^2 \left\langle |\tilde h_R|^2 + |\tilde h_L|^2 \right\rangle_s \\
        &\propto f^2 \left\langle |\tilde{h}^\gr_R|^2 e^{-2v_{\text{p}}(f)} + |\tilde{h}^\gr_L|^2 e^{2v_{\text{p}}(f)} \right\rangle_s.
\end{aligned}
\end{equation}
For isotropically-oriented sources in general relativity, $\langle |\tilde{h}^\gr_R|^2 \rangle_s = \langle |\tilde{h}^\gr_L|^2 \rangle_s$, and so we can write 
    \begin{equation}
    \left\langle\frac{dE_\ri}{df}+\frac{dE_\lef}{df}\right\rangle_s = \left\langle \frac{dE^\gr}{df} \right\rangle_s \cosh{2v_{\text{p}}(f)} ,
    \end{equation}
where we have defined 
\be 
\left \langle\frac{dE^\gr}{df} \right\rangle_s= \left \langle \frac{dE_\ri^\gr}{df} + \frac{dE_\lef^\gr}{df} \right\rangle_s
\ee 
as the net gravitational-wave energy per unit gravitational-wave frequency, radiated by a given source in general relativity.
Similarly, the difference between energies per frequency in right- and left-circular modes becomes
    \begin{equation}
    \left\langle\frac{dE_\ri}{df}-\frac{dE_\lef}{df}\right\rangle_s = \left\langle \frac{d E^\gr}{df} \right\rangle_s \sinh{2v_{\text{p}}(f)} .
    \end{equation}
Thus, one arrives at the modified expression for the Stokes-I and Stokes-V energy densities, namely 
\begin{align}
\Omega_I(f) = \frac{f}{H_0\rho_c}\int &dz \frac{R_m(z)}{(1 + z)\sqrt{\Omega_m(1 + z)^3 + \Omega_\Lambda}} \nonumber \\
&\times\left\langle\frac{d E^\gr}{df}\right\rangle_s \cosh{2v_{\text{p}}(f)} ,
\label{eq:stokesI}\\
\Omega_V(f) = \frac{f}{H_0\rho_c}\int &dz \frac{R_m(z)}{(1 + z)\sqrt{\Omega_m(1 + z)^3 + \Omega_\Lambda}} \nonumber \\
&\times\left\langle\frac{d E^\gr}{df}\right\rangle_s\sinh{2v_{\text{p}}(f)} .
\label{eq:stokesV}
\end{align}
We can see that both $\Omega_I(f)$ and $\Omega_V(f)$ are now exponentially enhanced due to the presence of the $\cosh 2v_{\text{p}}$ and $\sinh 2v_{\text{p}}$, respectively. Explicitly in terms of $\kappa_D$ and $\kappa_z$, the modification becomes
\begin{align}
    \Omega_I(f) & = \frac{f}{H_0\rho_c}\int dz \frac{R_m(z)}{(1 + z)\sqrt{\Omega_m(1 + z)^3 + \Omega_\Lambda}} \nonumber \\
&\times\left\langle\frac{d E^\gr}{df}\right\rangle_s \cosh\left[\frac{\pi f}{100 \text{Hz}}\left(\kappa_z z + \kappa_D\frac{D_C}{\text{Gpc}}\right)\right] ,
\label{eq:stokesI}\\
\Omega_V(f) &= \frac{f}{H_0\rho_c}\int dz \frac{R_m(z)}{(1 + z)\sqrt{\Omega_m(1 + z)^3 + \Omega_\Lambda}} \nonumber \\
&\times\left\langle\frac{d E^\gr}{df}\right\rangle_s\sinh\left[\frac{\pi f}{100 \text{Hz}}\left(\kappa_z z + \kappa_D\frac{D_C}{\text{Gpc}}\right)\right] .
\label{eq:stokesVkappa}
\end{align}
If one expands each of these in small $\kappa_D$ and $\kappa_z$, we recover the results from the theory-motivated approach, described next.



\subsubsection*{Approach 2: Theory-Motivated}
\label{sec:eft-amplification}


We now consider our more restricted, theoretically-motivated approach which is required to obey constraints imposed by effective field theory. This approach allows only small linear deviations from general relativity, but enables direct comparisons to known, parity-violating theories. The waveform modification is given by Eq.~\eqref{eq:hrlLinear}, with $v_{\text{th}}(f)$ given by Eq.~\eqref{eq:eft-v}.

In this case, the stochastic background I-mode contribution is modified only at $\mathcal{O}(v_{\text{th}}^2)$, such that we have~\cite{Yagi:2017zhb}
\be  
\Omega_I(f) = \Omega_I^\gr + \mathcal{O}(v_{\text{th}}^2).  
\label{eq:OmegaIth}
\ee  
One can see this explicitly by expanding Eq.~\eqref{eq:stokesI} in $v_{\text{th}}(f) \ll 1$.
Due to the smallness of the parity-violating coefficients in $v_{\text{th}}(f)$, any terms beyond linear order are negligible, and thus, for the theory motivated scenario, the I-mode is uninformative. 

When we consider the parity-sensitive V mode, however, we find that there is a linear-order correction, such that \cite{Yagi:2017zhb, Isi:2018miq}
\begin{align} 
\Omega_V(f) &\sim \frac{f}{H_0\rho_c}\int dz \frac{R_m(z)}{(1 + z)\sqrt{\Omega_m(1 + z)^3 + \Omega_\Lambda}}\nonumber \\
&\times\left\langle\frac{dE^\gr}{df}\right\rangle_s (2|v_{\text{th}}(f)|/c) \;  .
\label{eq:stokesV-linear}
\end{align} 
In terms of the parity-violating parameters $\alpha_1$ and $\beta_1$, we have 
\begin{align} 
\Omega_V(f) &\sim \frac{f}{H_0\rho_c}\int dz \frac{R_m(z)}{(1 + z)\sqrt{\Omega_m(1 + z)^3 + \Omega_\Lambda}}\nonumber \\
&\times\left\langle\frac{dE^\gr}{df}\right\rangle_s \left[2\pi f \left(\frac{\alpha_{1}}{\Lambda_{\pv}}z  + \beta_{1} D_C\right)\right] \;.
\label{eq:stokesV-linearalphabeta}
\end{align} 
From this, we see that while $\Omega_I(f)$ is unaffected, $\Omega_V(f)$ acquires a linear amplification that one can search for (or constrain) from the data. 

\begin{figure*}[htb]
    \centering
 \includegraphics[width=0.95\textwidth]{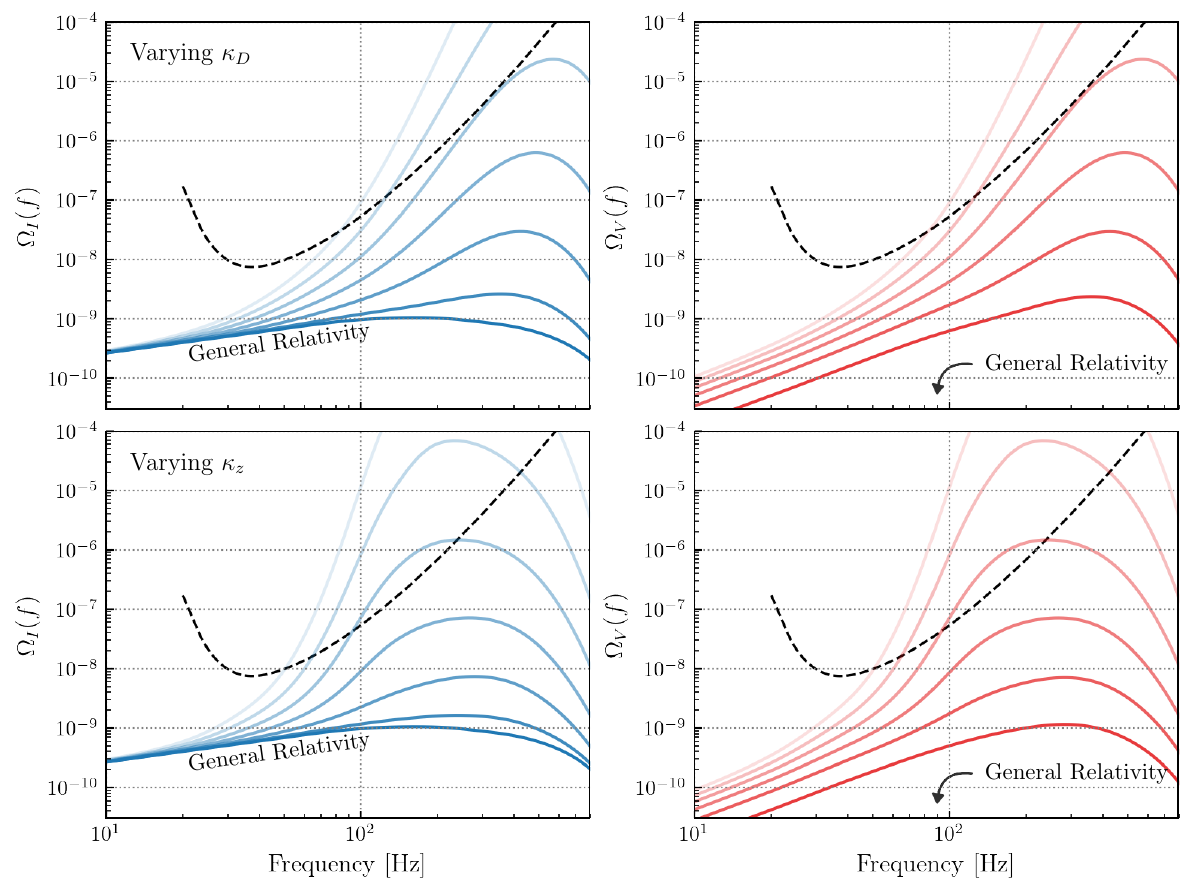}
    \caption{
    Illustration of the effect of amplitude birefringence on the energy-density of the gravitational-wave background.
    The left- and right-hand columns show the impacts on the spectra $\Omega_I(f)$ and $\Omega_V(f)$ of the Stokes I and V parameters, respectively.
    In the upper row we vary $\kappa_D$ while fixing $\kappa_z=0$, while in the bottom row we instead vary $\kappa_z$ while holding $\kappa_D=0$.
    In all four panels, the dashed black curve indicates the sensitivity of the most recent LIGO-Virgo search for the stochastic gravitational-wave background; spectra lying above these sensitivity curves are generally expected to have $\mathrm{SNR}\gtrsim 1$. Notice that increasing transparency in each curve corresponds to increasing birefringence.
    }
    \label{fig:birefringence_example}
\end{figure*}

\subsection{An illustrative example}
\label{sec:illustration}

We now illustrate how the parity-violation modifications to $\Omega_I$ and $\Omega_V$ discussed in the previous subsection can be used to place a constraint on birefringence from the \textit{non-detection} of a stochastic gravitational-wave background from compact binaries. 
Any mechanism that enhances the stochastic background (whether exponentially as in Eqs.~\eqref{eq:stokesI} and \eqref{eq:stokesV} or linearly as in Eq.~\eqref{eq:stokesV-linear}) must respect the fact that we do not observe such a background using present-day detectors~\cite{KAGRA:2021kbb}.
Since the compact binaries comprising the stochastic background occur at extreme distances~\cite{Callister:2016ewt}, they would undergo significant birefringent amplification, and thus the non-detection of the background can place strong limits on the degree of birefringence.

As a concrete example, let us focus on constraints on parity-violation from the stochastic background in the phenomenological scenario. 
Figure~\ref{fig:birefringence_example} illustrates the impact of amplitude birefringence on the energy density of the gravitational-wave background, where we have considered a binary black hole merger rate following low-metallicity star formation~\cite{Langer:2005hu,Madau:2014bja} subject to a distribution of evolutionary time delays (as we will discuss later, in Sec.~\ref{sec:methods}).
The top row corresponds to varying $\kappa_D$ while fixing $\kappa_z=0$, while the bottom panels conversely correspond to varying $\kappa_z$ while fixing $\kappa_D=0$.
The left column shows the resulting modifications to $\Omega_I(f)$; in both panels, the lowermost curve shows a standard prediction for an unamplified stochastic signal, while subsequent lighter curves correspond to increasing $\kappa_D$ or $\kappa_z$ (see also Fig.~\ref{fig:birefringence_simple_example} for a simplified version of the upper left panel).
The right column, meanwhile, shows the effect of birefringence on $\Omega_V(f)$.
In the absence of birefringence, $\Omega_V(f)$ is identically zero.
Increasing $\kappa_D$ or $\kappa_z$, however, gives non-zero and increasingly large $\Omega_V(f)$.
In the limit of extreme birefringence, $\Omega_V(f)$ asymptotes towards $\Omega_I(f)$.
This reflects the fact that, in the limit of extreme amplification, only a single polarization mode is non-zero, which implies that $|\Omega_V(f)| = \Omega_I(f)$.

The two ``types'' of birefringence discussed here (i.e., that due to $\kappa_z$ and that due to $\kappa_D$) impart different spectral shapes to the stochastic background. Large $\kappa_D$ yields a peak in $\Omega_I(f)$ centered at $f\sim 500\,\mathrm{Hz}$, while large $\kappa_z$ instead gives a lower-frequency peak at $f \sim 200\,\mathrm{Hz}$.
Qualitatively, we can understand this behavior from the competition of the two terms in Eq.~\eqref{eq:birefringent-v}.
On the one hand, birefringent amplification is strongest at large $f$.
On the other hand, amplification is also strongest for the most distant sources, whose gravitational-wave emission is redshifted towards \textit{lower} frequencies.
Birefringent amplification via $\kappa_z$ prefers higher redshift sources relative to $\kappa_D$, yielding a lower peak frequency in the gravitational-wave background.

In all four panels, the dashed black curve shows the ``power-law integrated'' (PI) curve illustrating the current sensitivity of the LIGO-Virgo detector network following the latest O3 run~\cite{Thrane:2013oya,KAGRA:2021kbb}.
By definition, a power-law energy-density spectrum lying tangent to the PI curve has expected signal-to-noise ratio $\langle\mathrm{SNR}\rangle=1$.
Although the birefringently amplified energy-density spectra do not have power-law forms, spectra with significant excursions above the PI curves are expected to be generally detectable, and so in principle, they can be ruled out. In order to determine the degree to which this is the case quantitatively, however, we must carry out a more careful data analysis, as we do in the next section. 

\section{Analysis Methods}
\label{sec:methods}

Searches for the stochastic gravitational-wave background rely on the cross-correlation of detector pairs.
Let $\tilde s_i(f)$ be the Fourier-transformed strain signal measured by detector $i$ (including contributions from both gravitational waves and instrumental noise).
Given two detectors, $i$ and $j$, we can define a cross-correlation statistic~\cite{Allen:1997ad,Romano:2016dpx}
\be
\hat C_{ij}(f) = \frac{1}{T} \frac{20 \pi^2}{3 H_0^2} f^3 \tilde s_i(f) \tilde s_j^*(f).
\label{eq:cross-corr-statistic}
\ee
This quantity is normalized such that its expectation value over an ensemble of noise realizations is
\be
\langle \hat C_{ij}(f) \rangle = \gamma_{ij}^R(f) \Omega_R(f) + \gamma_{ij}^L(f) \Omega_L(f)
\label{eq:cross-corr-circular}
\ee
and its variance is
\be
\langle \hat C_{ij}(f) \hat C_{ij}(f') \rangle = \delta(f-f') \sigma_{ij}^2(f),
\label{eq:variance}
\ee
with
\be
\sigma_{ij}^2(f) = \frac{1}{T} \left(\frac{10\pi^2}{3 H_0^2}\right)^2 f^6 P_i(f) P_j(f).
\label{eq:sigma-f}
\ee
In Eq.~\eqref{eq:sigma-f}, $T$ is the total observation time and $P_i(f)$ is the one-sided noise power-spectral density of detector $i$.

The quantities $\gamma_{ij}^R(f)$ and $\gamma_{ij}^L(f)$ in Eq.~\eqref{eq:cross-corr-circular} are the right- and left-circular \textit{overlap reduction functions}.
These functions quantify the sensitivity of a given detector pair to an isotropic background of each polarization, and depend on the geometry of the detector baseline.
Let $F_i^R(\hat n)$ and $F_i^L(\hat n)$ be the antenna patterns quantifying the response of a given detector to right- or left-circularly polarized gravitational waves arriving from direction $\hat n$.
In terms of the more standard antenna patterns $F_i^+(\hat n)$ and $F_i^\times(\hat n)$ for ``plus'' and ``cross'' polarizations~\cite{1987MNRAS.224..131S},
these are defined via
\be
F_i^R(\hat n) = F_i^+(\hat n) + i F_i^\times (\hat n)
\ee
and
\be
F_i^L(\hat n) = F_i^+(\hat n) - i F_i^\times (\hat n).
\ee
Then the right- and left-circular overlap reduction functions are
\be
\gamma^{R,L}_{ij}(f) = \frac{5}{8\pi} \int d\hat n\,
    F_i^{R,L}(\hat n) F_j^{R,L} (\hat n)^*
    e^{2\pi i f \Delta \mathbf{x}\cdot \hat n/c},
\label{eq:orfs-circular}
\ee
where $\Delta \mathbf{x}$ is the separation vector between detectors and $c$ is the speed of light.
The normalization of Eq.~\eqref{eq:orfs-circular} is chosen such that $\gamma^{R,L}_{ij}(f) = 1$ for coincident and colocated right-angle interferometers, like LIGO, Virgo, and KAGRA.

The expectation value of the cross-correlation statistic in the right-left polarization basis was presented in Eq.~\eqref{eq:cross-corr-circular}, but this quantity can equivalently be written in the Stokes I and V basis~\cite{Seto:2007tn},
    \be
    \langle \hat C_{ij}(f) \rangle = \gamma_{ij}^I(f) \Omega_I(f) + \gamma_{ij}^V(f) \Omega_V(f),
    \label{eq:cross-corr-stokes}
    \ee
with variance again given by Eq.~\eqref{eq:variance}.
Note that Eq.~\eqref{eq:cross-corr-stokes} depends on a new set of overlap reduction functions, $\gamma^I_{i,j}(f)$ and $\gamma^V_{i,j}(f)$.
Rewriting Eq.~\eqref{eq:cross-corr-circular} in terms of $\Omega_I(f)$ and $\Omega_V(f)$ with Eqs.~\eqref{eq:OmgI} and~\eqref{eq:OmgV}, and comparing to Eq.~\eqref{eq:cross-corr-stokes}, we find
\be
\gamma^I_{ij}(f) = \gamma^R_{ij}(f) + \gamma^L_{ij}(f)
\ee
and
\be
\gamma^V_{ij}(f) = \gamma^R_{ij}(f) - \gamma^L_{ij}(f).
\ee
In practice, $\gamma^I_{ij}(f)$ and $\gamma^V_{ij}(f)$ can be directly and analytically expressed as linear combinations of spherical Bessel functions~\cite{Flanagan:1993ix,Seto:2007tn}

Let $\Omega_I^M(f;\Lambda)$ and $\Omega_V^M(f;\Lambda)$ be models for the expected stochastic gravitational-wave background, given some set of parameters $\Lambda$; this set includes $\kappa_D$ and $\kappa_z$, but might also include other parameters governing the strength of the stochastic background (such as parameters describing the rate of black hole mergers as a function of redshift, as in Sec.~\ref{sec:variable-rate}).
Then, the likelihood of measuring cross-correlation $\hat C_{ij}(f)$ between detectors $i$ and $j$ is
\be
p(\hat C_{ij}\,|\,\Lambda) \propto
    \mathrm{exp}\left[ - \frac{1}{2} \left(\hat C_{ij} - \gamma^A_{ij}\Omega_A^M | \hat C_{ij} - \gamma^A_{ij} \Omega_A^M\right)\right],
    \label{eq:likelihood-individual}
\ee
assuming Gaussianity of instrumental noise.\footnote{In practice, $\hat C_{ij}$ is estimated by combining a large number of independent cross-correlation spectra measured over a large number of $\mathcal{O}(10^2\,\mathrm{s})$ intervals, in which case Gaussianity is also ensured by the central limit theorem.}
Here, we abbreviate $\gamma^A_{ij}\Omega_A^M = \gamma^I_{ij} \Omega_I^M + \gamma^V_{ij} \Omega_V^M$ and have defined the inner product
\be
(A|B) = \int_0^\infty df \frac{A(f) B^*(f) + A^*(f) B(f)}{\sigma^2_{ij}(f)}.
\label{eq:inner-product}
\ee
In the limit that the gravitational-wave background is much weaker than instrumental noise, cross-correlation measurements between different baselines are independent (even if those baselines share detectors in common).
The full likelihood for multiple sets of cross-correlation measurements can therefore be factorized as a product over each individual detector pair:
\be
p(\{\hat C_{ij}\}\,|\,\Lambda) = \prod_{{\rm Pairs\,}ij} p(\hat C_{ij}\,|\,\Lambda).
\label{eq:likelihood-full}
\ee

\begin{figure*}[htb!]
    \centering
    \includegraphics[width=0.95\textwidth]{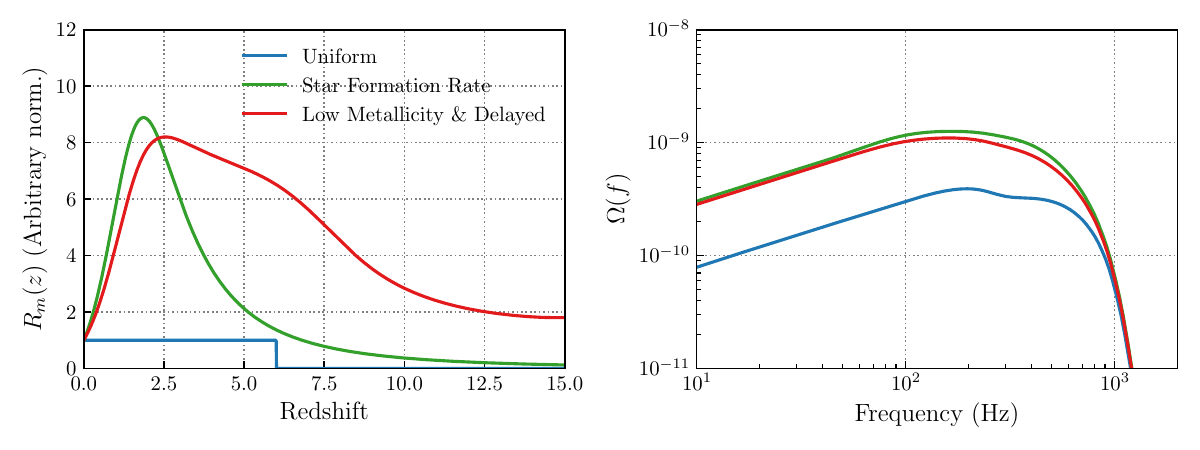}
    \caption{
    \textit{Left:}
    Three source-frame binary black hole merger rate models considered in this work.
   Constraints on birefringent amplification of the gravitational-wave background are degenerate with assumptions regarding the merger history of binary black holes.
    We bracket this systematic uncertainty by considering the three models shown: a constant uniform-in-comoving volume rate (blue), a merger rate directly tracing the global star formation rate (green), and a merger rate tracing low-metallicity star formation subject to a distribution of evolutionary time delays (red).
    \textit{Right:} The stochastic energy-density spectra predicted under each model.
    With the lowest integrated merger rate, the uniform-in-comoving volume model (blue) provides a deliberate underestimate of the gravitational-wave background; any constraints derived from this model will therefore be conservative.
    The remaining models, each more astrophysically realistic, predict larger $\Omega(f)$ and hence more optimistic constraints on gravitational-wave birefringence.
    The most optimistic constraints arise when using the delayed low-metallicity rate model (red).
    Despite the fact that the stochastic background under this model is not the strongest of the three, this model places sources at the largest redshifts and hence maximizes the expected degree of amplitude birefringence.
    }
    \label{fig:rate_comparison}
\end{figure*}

To calculate the expected stochastic gravitational-wave background and compare it against data, we also need to know the properties of the compact binaries contributing to the background.
Recall that the stochastic background depends on the ensemble-averaged energy spectra of binary sources:
    \begin{equation}
    \begin{aligned}
    \left\langle\frac{dE}{df}\right\rangle_s
        &= \int dm_1\,dm_2\,d\vec\chi_1\,d\vec\chi_2 
        \frac{dE}{df}(m_1,m_2,\vec\chi_1,\vec\chi_2) \\
        & \hspace{5mm}\times p(m_1) p(m_2|m_1) p(\vec\chi_1,\vec\chi_2)
    \end{aligned}
    \end{equation}
where $\vec\chi_1$ and $\vec\chi_2$ are the spin vectors of the component black holes, and $p(m_1)$, $p(m_2|m_1)$, and $p(\vec\chi_1,\vec\chi_2)$ are the probability distributions of primary masses, secondary masses (given primary masses), and component spins characterizing the binary black hole population.
We describe the black hole mass distribution as a mixture between a broad power-law and a Gaussian peak centered near $35\,M_\odot$, consistent with the latest observational results~\cite{KAGRA:2021duu}; this same model was used to calculate Figs.~\ref{fig:birefringence_simple_example} and \ref{fig:birefringence_example} above.
More details regarding the exact mass model used are shown in Appendix~\ref{app:MassDist}.
Black hole spins, meanwhile, are known from population inferences to be somewhat small but not identically zero~\cite{KAGRA:2021duu,Callister:2022qwb,Mould:2022xeu,Callister:2023tgi}.
The effect of non-zero spin on the gravitational-wave background is small~\cite{Zhu:2011bd,LIGOScientific:2016fpe}, however, and so for simplicity we assume black hole component spin magnitudes of zero.
We evaluate the energy spectra themselves using the inspiral-merger-ringdown model of~\cite{Ajith:2009bn}, including the given post-Newtonian corrections.

We also need a prescription for the merger rate history of binary black holes.
The expected stochastic background amplitude (and hence our constraints on birefringence) will depend strongly on one's exact choice of $R_m(z)$.
The local rate of merging black holes near redshift $z=0$ is reasonably well known; unless stated otherwise we will assume $R_0 \equiv R_m(0) = 16\,\mathrm{Gpc}^{-3}\,\mathrm{yr}^{-1}$, consistent with estimates from Ref.~\cite{KAGRA:2021duu}.\footnote{Note that the merger rates appearing in Table~IV of Ref.~\cite{KAGRA:2021duu} are those at $z=0.2$ rather than $z=0.0$. Since the merger rate is known to rise with redshift, these rates are accordingly larger than the rate adopted here.}
The evolution of the merger rate towards higher redshift, however, remains highly uncertain~\cite{Fishbach:2018edt,Callister:2020arv,KAGRA:2021duu,Chruslinska:2022ovf,Edelman:2022ydv,Callister:2023tgi,Fishbach:2023pqs}.
We therefore proceed by adopting several different estimates for $R_m(z)$, shown in Fig.~\ref{fig:rate_comparison}, in order to bracket this systematic uncertainty.

\vspace{0.2cm}
\noindent \textit{1. Uniform-in-comoving volume.}
One choice is to simply assume a constant, uniform-in-comoving volume merger rate out to some maximum redshift ($z\approx 6$) beyond which we expect minimal star formation:
    \be
    R_m^{\rm const}(z) =
        \begin{cases}
        R_0 & (z\leq 6) \\
        0 & (\rm z>6)\,.
        \end{cases}
    \label{eq:rate-model-constant}
    \ee
This model is shown in blue on the left panel of Fig.~\ref{fig:rate_comparison}.
The right panel of Fig.~\ref{fig:rate_comparison}, meanwhile, shows in blue the energy-density spectrum, $\Omega_I(f)$, predicted under this model in GR. 
Since binary black hole merger rate is known to \textit{increase} with redshift~\cite{KAGRA:2021duu}, rather than remain constant, this model is almost certain to systematically underestimate the total integrated merger rate.
It should correspondingly underestimate the stochastic gravitational-wave background.
This implies that any bounds on birefringence obtained under this choice of $R_m(z)$ will be \textit{conservative}: by underestimating the gravitational-wave background, we maximize the degree of birefringent amplification that remains consistent with a non-detection.

\vspace{0.2cm}
\noindent \textit{2. Directly tracing global star formation.}
If binary black holes arise from stellar progenitors, then we generically expect the merger rate to evolve similarly to the Universe's global star formation history, rising into the past, reaching a peak near ``cosmic noon'' at $z\approx 2$, and subsequently decaying at very large redshift.
For our second model, we take $R_m(z)$ to be directly proportional to the global star formation rate inferred in~\cite{Madau:2014bja}:
    \be
    R_m^\sfr(z) = \frac{R_0}{\mathcal{C}}\frac{(1+z)^\alpha}{1+\left(\frac{1+z}{1+z_p}\right)^{\alpha+\beta}}
    \label{eq:rate-model-sfr}
    \ee
with $\alpha = 2.7$, $\beta = 2.9$, and $z_p = 1.9$.
We truncate the merger rate at $z_\mathrm{max} = 15$, and assume it is identically zero at higher redshifts.
The normalization constant
    \be
    \mathcal{C} = \left[1 + \frac{1}{(1+z_p)^{\alpha+\beta}}\right]^{-1}
    \ee
ensures that $R_m^\sfr(z) = R_0$ at $z=0$. 
Adopting Eq.~\eqref{eq:rate-model-sfr} yields the green curves in Fig.~\ref{fig:rate_comparison}.
Relative to the uniform-in-comoving volume model, $R_m^\sfr(z)$ predicts many more binary black hole mergers at large redshifts, yielding an elevated stochastic gravitational-wave background.

\vspace{0.2cm}
\noindent \textit{3. Delayed tracer of low-metallicity star formation.}
Black hole mergers are unlikely to be direct tracers of the global star formation rate for two reasons.
First, black holes are expected to form more efficiently in low-metallicity environments~\cite{Chruslinska:2018hrb,Giacobbo:2018etu,Chruslinska:2022ovf}.
Second, black holes are likely to experience a range of evolutionary time delays between the births of their stellar progenitors and their eventual binary mergers~\cite{Giacobbo:2018etu,Neijssel:2019irh,vanSon:2021zpk,Fishbach:2023pqs}.
Our final and most realistic model seeks to capture these features.
We assume that the rate of binary black hole \textit{births} is
    \be
    R_{\rm birth}(z) = R_m^\sfr(z)\,F(Z<Z_\mathrm{thresh},z),
    \ee
following the same function as in Eq.~\eqref{eq:rate-model-sfr} but now weighted by the cumulative fraction $F(Z<Z_\mathrm{thresh},z)$ of star formation occurring at metallicities $Z$ below a cutoff metallicity $Z_\mathrm{thresh}$.
We take $Z_\mathrm{thresh} = 0.1 Z_\odot$~\cite{Giacobbo:2018etu,Chruslinska:2018hrb,Fishbach:2023pqs} and calculate $F(Z<Z_\mathrm{thresh})$ following the fitting formula of Ref.~\cite{Langer:2005hu}.
The \textit{merger} rate of binaries is then obtained through the convolution of $R_{\rm birth}(z)$ with a probability distribution of evolutionary time delays $t_d$:
    \be
    R_m^{\rm delayed}(z) = \int dt_d\,R_{\rm birth}(z_b(z,t_d))\,p(t_d).
    \label{eq:rate-model-delayed}
    \ee
Here, we use $z_b$ to indicate birth redshift and write $z_b \equiv z_b(z,t_d)$ to indicate that it should be regarded as a function of time delay and eventual merger redshift $z$.
We assume that time delays are distributed log-uniformly, with $p(t_d) \propto t_d^{-1}$ for $10\,{\rm Myr}<t_d<13.5\,{\rm Gyr}$, and zero otherwise.
The merger rate density and energy-density spectrum given by this final model are shown in red in Fig.~\ref{fig:rate_comparison}.
We see that this model places sources at even larger redshifts than the star-formation-tracing model above.
The standard energy-density spectrum predicted by $R_m^{\rm delayed}(z)$ is actually slightly reduced relative to the prediction from $R_m^\sfr(z)$.
However, because it places sources at such large redshifts, $R_m^{\rm delayed}(z)$ leaves greater opportunities for sigificant birefringent amplification, particularly when $\kappa_z$ is nonzero.
We will see, therefore, that this final ``most realistic'' model of the black hole merger rate gives the most optimistic constraints on birefringence.
This is the model used to generate the spectra shown in Fig.~\ref{fig:birefringence_example}.

With the above models for the mass and redshift distributions of binary black hole mergers, we constrain the degree of allowed birefringence using cross-correlation spectra measured between the LIGO-Hanford, LIGO-Livingston, and Virgo detectors.
Specifically, we analyze the Hanford-Livingston cross-correlation spectra in their O1, O2, and O3 observing runs, and the Hanford-Virgo and Hanford-Livingston cross-correlation spectra during O3.\footnote{https://dcc.ligo.org/G2001287/public}
Note that our likelihood, as defined in Eqs.~\eqref{eq:likelihood-individual} and \eqref{eq:likelihood-full}, does not account for imperfect detector calibration.
Explicitly marginalizing over amplitude calibration uncertainty would yield percent-level shifts to the results shown below~\cite{Yousuf:2023nmz,Renzini:2023qtj}.

\section{Gravitational Wave Constraints on Phenomenological Birefringence}
\label{sec:phenomenological-results}

As described in Sec.~\ref{sec:parity-violation}, we can approach gravitational-wave amplitude birefringence in two ways: purely phenomenologically, or as a small perturbation from general relativity.
We begin here with the former, in which we allow birefringence to inflate the gravitational-wave background arbitrarily without restriction.
Although this method does not allow us to compare directly to known parity-violating theories, it does allow us to place a meaningful theory-independent bound on gravitational parity violation based on the present non-detection of a stochastic gravitational-wave background.

\subsection{Fixed Merger Rate}
\label{sec:fixed-rate}

\begin{figure*}[htb]
    \centering
    \includegraphics[width=0.7\textwidth]{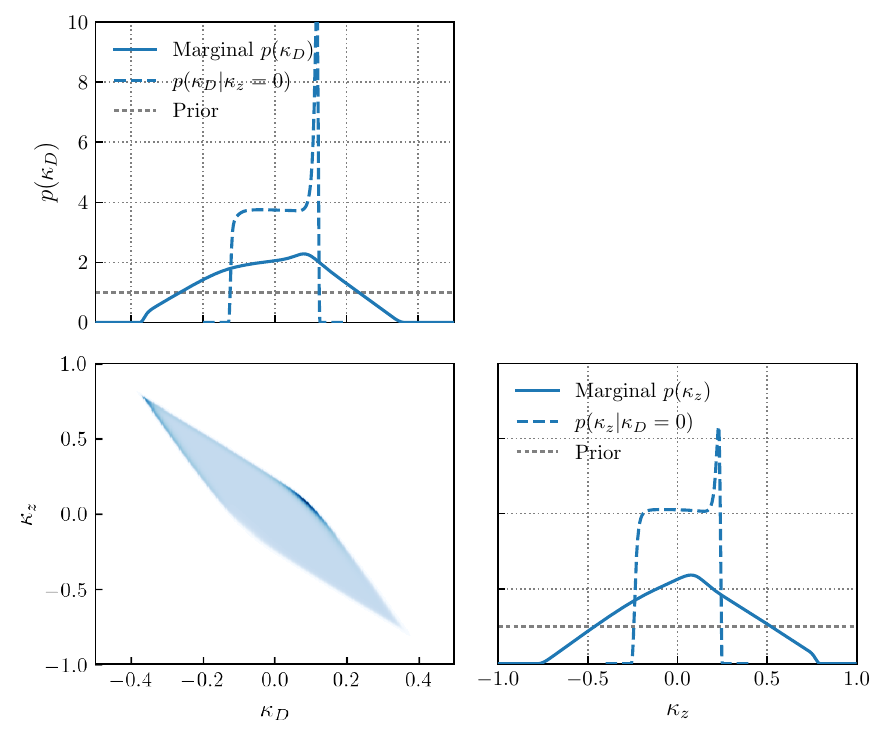}
    \caption{ 
    Joint posterior on the coefficients $\kappa_D$ and $\kappa_z$ governing the degree of amplitude birefringence as a function of comoving distance and redshift, respectively (see Eq.~\eqref{eq:birefringent-v}), when assuming a binary black hole merger rate that is constant in comoving volume (Eq.~\eqref{eq:rate-model-constant}).
    These constraints are obtained under the ``phenomenological'' model described in Sec.~\ref{sec:phenom-amplification}, in which birefringent amplification is not constrained to linear order but allowed to grow arbitrarily large.
    The panels along the diagonal show the corresponding one-dimensional posteriors on $\kappa_D$ and $\kappa_z$; solid curves show the marginalized posterior on each coefficient, while dotted curves show conditional posteriors obtained when the other parameter is fixed to zero.
    The absence of a stochastic background  detection places the marginalized limits $\kappa_D = \UniformRateKappaDcMarginal$ and $\kappa_z = \UniformRateKappaZMarginal$ on the birefringent coefficients, or the conditional limits $\kappa_D = \UniformRateKappaDcConditioned$ and $\kappa_z = \UniformRateKappaZConditioned$ when the other is taken to zero.
    As a constant-in-comoving-volume merger rate almost certainly underestimates the gravitational-wave background, greater degrees of birefringence are permitted by a non-detection of that background.
    These constraints can accordingly be taken as deliberately conservative; more realistic constraints are shown below in Fig.~\ref{fig:SFR_results}.
    Although both birefringent coefficients are consistent with zero, we do see a narrow edge of elevated probability in their joint posterior, manifesting as narrow ``spikes'' in the conditional $\kappa_D$ and $\kappa_z$ posteriors.
    This is explored further in Appendix~\ref{app:spike}, where we trace this feature to marginally-significant excess narrowband cross-power in the Hanford-Livingston and Hanford-Virgo baselines.
    }
    \label{fig:uniform_rate_results}
\end{figure*}

\begin{figure*}
    \centering
    \includegraphics[width=0.98\textwidth]{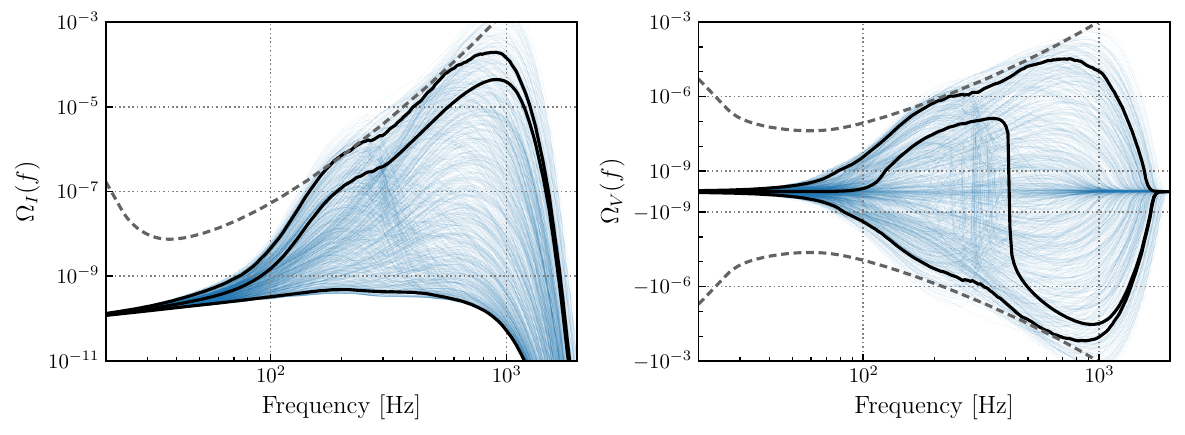}
    \caption{
    Limits on the Stokes-I (left) and Stokes-V (right) energy densities of the stochastic gravitational-wave background from binary black holes in the presence of parity-violating birefringent amplification.
    Each trace represents a single random draw from the posterior on birefringent coefficients $\kappa_D$ and $\kappa_z$ shown in Fig.~\ref{fig:uniform_rate_omega}.
    The outermost black lines in each panel mark the central 90\% bounds on each spectrum, while the innermost black curves trace the posterior means.
    The dashed curves, called ``power-law integrated'' (PI) curves, indicate the most recent search sensitivities of the LIGO-Virgo detector network.
    Power-law spectra lying above (left panel) or outside (right panel) the PI curves are expected to be marginally detectable with $\mathrm{SNR}\geq 1$.
    The non-detection of a stochastic background therefore limits $\Omega_I(f)$ and $\Omega_V(f)$ to largely lie below or inside their respective PI curves.
    Note that, for these results, we have assumed that the black hole merger rate is uniform in comoving volume and vanishes above $z=6$ (see Eq.~\eqref{eq:rate-model-constant}).
    This choice is made to deliberately underestimate the stochastic gravitational-wave background, in order to obtain conservative bounds on $\kappa_D$ and $\kappa_z$ (see Fig.~\ref{fig:uniform_rate_results}).
    A series of increasingly more realistic models for the black hole merger rate are considered below.
    Although these subsequent models yield progressively tighter bounds on amplitude birefringence, their posteriors on $\Omega_I(f)$ and $\Omega_V(f)$ are extremely similar to those shown here.
    }
    \label{fig:uniform_rate_omega}
\end{figure*}

\begin{figure*}[htb]
    \centering
    \includegraphics[width=0.495\textwidth]{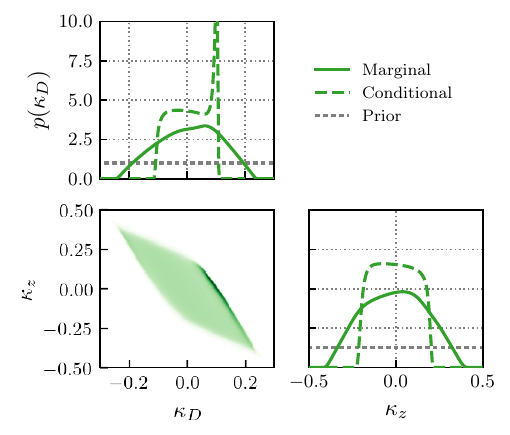}
     \includegraphics[width=0.495\textwidth]{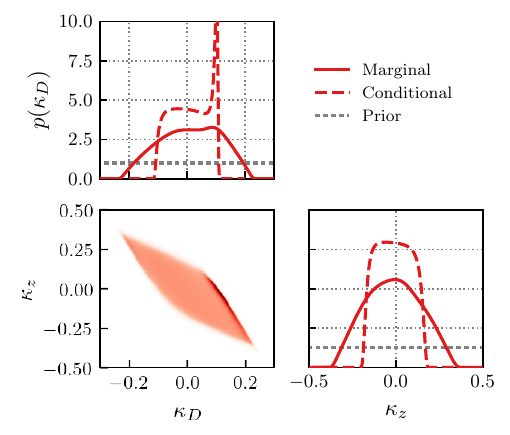}
    \caption{ 
    As in Fig.~\ref{fig:uniform_rate_results}, but assuming a binary black hole merger rate directly tracing global star formation (see Eq.~\eqref{eq:rate-model-sfr}) in the left corner plot, while assuming a rate that traces low metallicity star formation, subject to a distribution of evolutionary time delays (see Eq.~\eqref{eq:rate-model-delayed}), in the right corner plot.
    These more realistic models predict a larger gravitational-wave background, and the non-detection of this background accordingly gives stronger constraints on possible birefringent amplification.
    When the rate traces star formation, we obtain the limits $\kappa_D = \SfrKappaDcMarginal$ and $\kappa_z = \SfrKappaZMarginal$ on each coefficient when marginalizing over the other, or $\kappa_D = \SfrKappaDcConditioned$ and $\kappa_z = \SfrKappaZConditioned$ when conditioning the other to zero.
    When the rate adjusts for low-metallicities and delays, we obtain the limits $\kappa_D = \DelayedSfrKappaDcMarginal$ and $\kappa_z = \DelayedSfrKappaZMarginal$ on each coefficient when marginalizing over the other, or $\kappa_D = \DelayedSfrKappaDcConditioned$ and $\kappa_z = \DelayedSfrKappaZConditioned$ when conditioning the other to zero. 
    }
    \label{fig:SFR_results}
\end{figure*}

We consider, in turn, each of the three models described in Sec.~\ref{sec:methods} for the rate evolution of binary black hole mergers.
We fix the presumed mass and spin distributions of black holes to be consistent with latest measurements of the compact binary population~\cite{KAGRA:2021duu}; see Appendix~\ref{app:MassDist} for a full description.
The birefringent coefficients $\kappa_D$ and $\kappa_z$ (defined in Eq.~\eqref{eq:birefringent-v}) therefore remain the only two unknown parameters in our model.
We adopt uniform priors for both coefficients and, given the low dimensionality of our parameter space, directly compute likelihoods over a grid of possible parameter values.

Figure~\ref{fig:uniform_rate_results} shows the joint constraint on the birefringent parameters $\kappa_D$ and $\kappa_z$ when assuming that the black hole merger rate remains constant out to $z=6$.
The lower-left panel gives the joint posterior on both coefficients, while solid curves in the upper and lower-right panels give marginalized one-dimensional posteriors on each parameter.
We find that the non-detection of the stochastic gravitational-wave background implies $\kappa_D = \UniformRateKappaDcMarginal$ and $\kappa_z = \UniformRateKappaZMarginal$ for the median and 90\% confidence values.
The dashed curves, meanwhile, show the conditional posteriors on $\kappa_D$ and $\kappa_z$ obtained when the other parameter is fixed to zero.
This yields more restrictive limits, requiring $\kappa_D = \UniformRateKappaDcConditioned$ and $\kappa_z = \UniformRateKappaZConditioned$.
As discussed in Sec.~\ref{sec:methods}, assuming a purely constant merger rate almost certainly underestimates the stochastic gravitational-wave background, and so correspondingly \textit{overestimates} the degree of birefringence permitted by a non-detection of the stochastic background.
Thus these constraints are conservative.

There do exist several notable features in Fig.~\ref{fig:uniform_rate_results}.
Within the joint $\kappa_D\mbox{--}\kappa_z$ posterior we see that some quadrants (upper right and lower left) are constrained much more tightly than others (upper left and lower right).
When $\kappa_D$ and $\kappa_z$ share the same sign (upper right and lower left quadrants), they act in unison, each working to amplify the same polarization mode and yielding a stronger stochastic signal.
The \textit{non-detection} of such a signal, therefore, corresponds to relatively stringent constraints on these parameters.
When $\kappa_D$ and $\kappa_z$ have opposite signs (upper left and lower right quadrants), they instead work to counteract one another, each suppressing the polarization mode that the other seeks to amplify, and thus limiting the net amplification of the gravitational-wave background.
This cancellation is imperfect, as $\kappa_z$ and $\kappa_D$ impart different distance dependence for amplification (and hence distinct frequency dependence in the stochastic energy-density spectrum as in Fig.~\ref{fig:birefringence_example}), but our constraints are nevertheless weaker in these cases.

Although our $\kappa_D\mbox{--}\kappa_z$ posterior is consistent with zero, there is a very narrow feature with elevated posterior probability (see the dark ``edge'' on the right side of the two-dimensional posterior) centered around $\kappa_D \approx 0.15$ and $\kappa_z \approx 0$.
This band of elevated probability is more clearly seen in the conditional posteriors $p(\kappa_D|\kappa_z=0)$ and $p(\kappa_z|\kappa_D=0)$ (dotted curves), where it appears as a set of sharp ``spikes.''
Although this feature is not statistically significant, it is perhaps indicative of elevated cross-correlation between one or more pairs of detectors, which can arise from unlucky noise realizations or unidentified instrumental artifacts.
We will investigate this feature in further detail in Appendix~\ref{app:spike}.

In Fig.~\ref{fig:uniform_rate_omega} we translate our posteriors on $\kappa_D$ and $\kappa_z$ into measured constraints on Stokes-I and Stokes-V energy density spectra.
Recall that, while $\Omega_I(f)$ is positive-valued, $\Omega_V(f)$ can be either positive or negative depending on whether right- or left-circular polarizations dominate.
In the left-hand panel, the dashed line indicates the ``power-law integrated'' (PI) curve~\cite{Thrane:2013oya} denoting the sensitivity of the Hanford-Virgo-Livingston network following their most recent observing run~\cite{KAGRA:2021kbb}.
The PI curve is defined such that power-law spectra lying tangent to it are expected to have $\mathrm{SNR}=1$.
Correspondingly, we see that $\Omega_I(f)$ is constrained to lie below the PI curve, consistent with a non-detection of the stochastic background.
The dashed curves on the right-hand panel show the equivalent PI curves for Stokes-V energy densities.
As $\Omega_V(f)$ can be both positive and negative, two PI curves are shown, one for positive and one for negative amplitudes.
The non-observation of a gravitational-wave background constrains $\Omega_V(f)$ to lie between both PI curves.

\begin{table*}
    \setlength{\tabcolsep}{6pt}
    \renewcommand{\arraystretch}{1.3}
    \centering
    \caption{
    Constraints on birefringence coefficients $\kappa_D$ and $\kappa_z$ (defined in Eq.~\eqref{eq:birefringent-v}) under different models for the binary black hole merger rate $R(z)$.
    The first three rows correspond to the three different fixed choices for $R(z)$.
    As discussed in Secs.~\ref{sec:methods} and \ref{sec:fixed-rate}, these three models are progressively more realistic: the first (``Uniform'') deliberately underestimates the expected gravitational-wave background and thus provides the weakest bounds on amplitude birefringence, whereas the third (``Delayed star formation rate'') is most realistic and provides the tightest constraints.
    We also show constraints obtained in Sec.~\ref{sec:variable-rate}, when we do not fix the black hole merger but instead \textit{infer} it alongside $\kappa_D$ and $\kappa_z$ and marginalize over our uncertainty in $R(z)$.
    The numbers quoted correspond to medians and central 90\% credible intervals.
    We furthermore give both the marginal constraints on each parameter, as well as the conditional constraints obtained when fixing the other parameter to zero.
    }
    \begin{tabular}{r | r r r r}
    \hline \hline
    Merger Rate & $\kappa_D$ (marginal) & $\kappa_D$ ($\kappa_z=0$) & $\kappa_z$ (marginal) & $\kappa_z$ ($\kappa_D=0$) \\
    \hline
    Uniform; Eq.~\eqref{eq:rate-model-constant}
        & $\UniformRateKappaDcMarginal$
        & $\UniformRateKappaDcConditioned$
        & $\UniformRateKappaZMarginal$
        & $\UniformRateKappaZConditioned$ \\
    Star Formation Rate; Eq.~\eqref{eq:rate-model-sfr}
        & $\SfrKappaDcMarginal$
        & $\SfrKappaDcConditioned$
        & $\SfrKappaZMarginal$
        & $\SfrKappaZConditioned$ \\
    Delayed Star Formation Rate; Eq.~\eqref{eq:rate-model-delayed}
        & $\DelayedSfrKappaDcMarginal$
        & $\DelayedSfrKappaDcConditioned$
        & $\DelayedSfrKappaZMarginal$
        & $\DelayedSfrKappaZConditioned$ \\
    \hline
    Unknown Merger Rate
        & $\VariableRateKappaDcMarginal$
        & $\VariableRateKappaDcConditioned$
        & $\VariableRateKappaZMarginal$
        & $\VariableRateKappaZConditioned$ \\
    \hline
    \hline
    \end{tabular}
    \label{tab:kappa-constraints}
\end{table*}

Figure~\ref{fig:SFR_results}, in turn, shows our posterior on $\kappa_D$ and $\kappa_z$ when we adopt our more accurate models for the black hole merger rate: $R_m^{\rm SFR}(z)$ and $R_m^{\rm delayed}(z)$ (Eqs.~\eqref{eq:rate-model-sfr} and \eqref{eq:rate-model-delayed}, respectively).
We find the same qualitative features as in Fig.~\ref{fig:uniform_rate_results}: an elongated diagonal posterior with a narrow high-probability ``edge'' at positive $\kappa_D$.
Both $R_m^{\rm SFR}(z)$ and $R_m^{\rm delayed}(z)$, however, predict a larger stochastic background than the uniform-in-comoving-volume model above, and so result in tighter constraints on $\kappa_D$ and $\kappa_z$.
We list these constraints in Table~\ref{tab:kappa-constraints}, both when marginalizing over the joint $\kappa_D$--$\kappa_z$ distribution, as well as when instead requiring only one coefficient to be non-zero at a time.
The best constraints on $\kappa_z$ are obtained under $R_m^{\rm delayed}(z)$, which places binary black holes at the largest redshifts, and hence, maximizes the allowed degree of amplification.
At the same time, it is interesting to note that the constraints on $\kappa_D$ under this model are effectively identical to those obtained with $R_m^{\rm SFR}(z)$.
This is because, in the sufficiently distant Universe, moving binaries to higher redshifts does not appreciably change their \textit{comoving distances} (moving from $z=6$ to $z=9$ increases comoving distance only from $8.4$ to $9.4\,\mathrm{Gpc}$, for example).
Thus, although the redshift distributions produced by both models are very different, their distance distributions are similar.

\begin{figure*}[htb]
    \centering
    \includegraphics[width=0.98\textwidth]{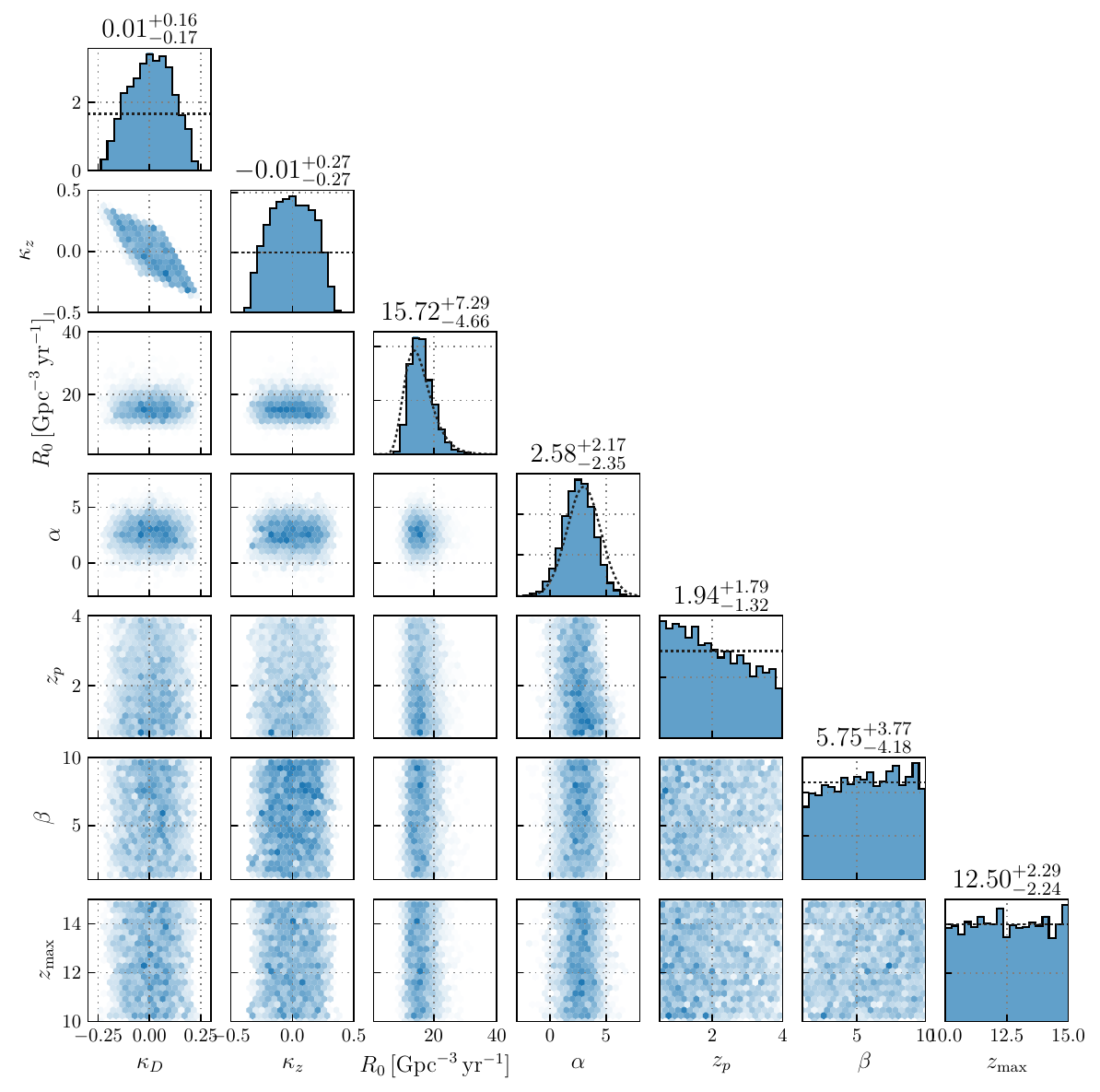}
    \caption{
    Full posteriors when simultaneously inferring birefringence coefficients alongside the binary black hole merger rate.
    The first two columns show the posteriors on coefficients $\kappa_D$ and $\kappa_z$, and the remaining columns correspond to parameters governing the height and shape of the binary black hole merger rate as a function of redshift; see Eq.~\eqref{eq:rate-model-sfr}.
    Within each marginalized posterior, dotted curves show the prior on each parameter.
    When marginalizing uncertainties in the black hole merger rate, we find constraints on $\kappa_D$ and $\kappa_z$ comparable to those obtained under fixed merger rate models above.
    Exact constraints are listed in Table~\ref{tab:kappa-constraints}.
    }
    \label{fig:full_corner}
\end{figure*}

\begin{figure}
    \centering
    \includegraphics[width=0.46\textwidth]{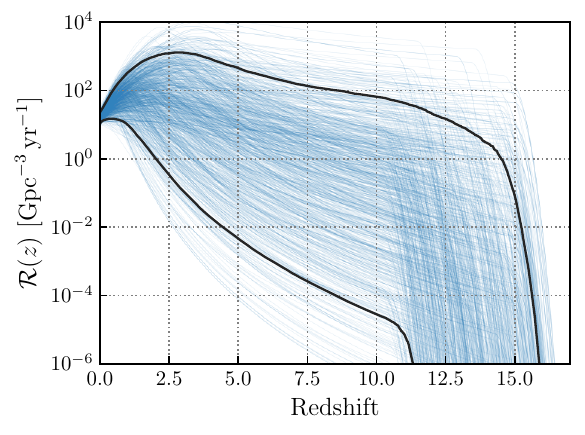}
    \caption{
    Uncertainty on the merger rate of binary black holes as a function of redshift, obtained when simultaneously fitting for gravitational-wave birefringence as well as the shape of the black hole merger rate.
    Each light trace represents a single posterior sample drawn from Fig.~\ref{fig:full_corner}.
    The thin outermost black lines denote the 90\% credible bounds on the merger rate, while the inner thick black line traces the posterior mean as a function of redshift.
    }
    \label{fig:rate_constraints}
\end{figure}

\begin{figure*}
    \centering
    \includegraphics[width=0.98\textwidth]{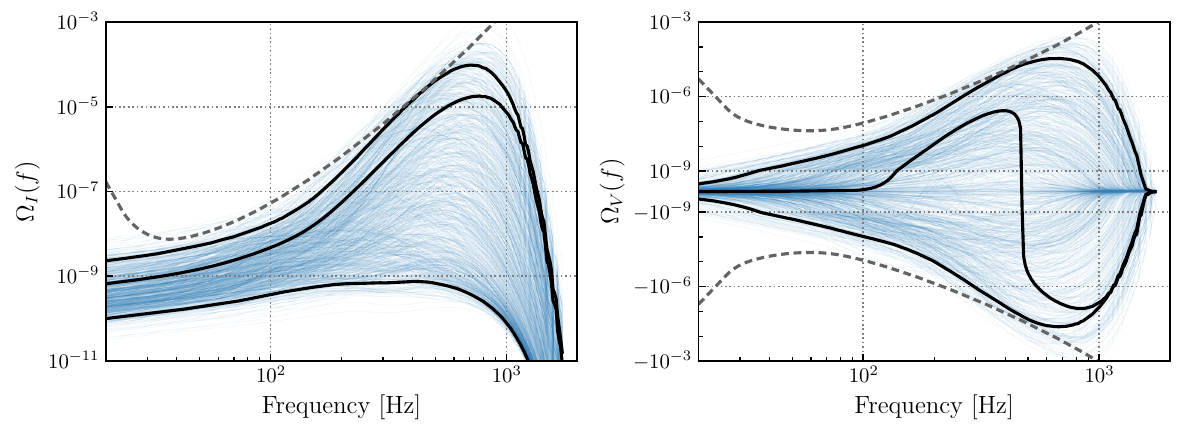}
    \caption{ 
    Constraints on the Stokes-I (left) and Stokes-V (right) energy-density spectra of the stochastic background, obtained when fitting for both the degree of gravitational-wave birefringence as well as the binary black hole merger history (compare to Fig.~\ref{fig:uniform_rate_omega} in which the merger rate is held fixed).
    Each trace corresponds to a single draw from the posterior in Fig.~\ref{fig:full_corner}.
    The outermost black curves in each panel mark the central 90\% credible bounds on $\Omega_I(f)$ and $\Omega_V(f)$, while the inner black curve gives the mean posterior value of each spectrum.
    The dashed grey lines show PI curves depicting the sensitivity of the Hanford-Livingston-Virgo network following their latest O3 observing run.
    The non-detection of a gravitational-wave background requires the gravitational-wave energy density to fall below (left) or inside (right) these PI curves.
    }
    \label{fig:OmgGW_constraints}
\end{figure*}

\subsection{Unknown Merger Rate}
\label{sec:variable-rate}

As illustrated by results in Sec.~\ref{sec:fixed-rate}, the largest source of systematic uncertainty in this work is our incomplete knowledge of the merger rate history of binary black holes.
Different presumed merger rate histories vary the anticipated gravitational-wave background by factors of several (see Fig.~\ref{fig:rate_comparison}), and therefore, alter the degree to which a non-detection of this background rules out birefringent amplification.
In Sec.~\ref{sec:fixed-rate}, we dealt with this by attempting to bracket the relevant uncertainty, providing results under both realistic (Fig.~\ref{fig:SFR_results}) and deliberately over-conservative (Fig.~\ref{fig:uniform_rate_results}) assumptions regarding the black hole merger rate.
An alternative approach, investigated here, is to relax our assumptions further and instead \textit{fit} for the binary black hole merger alongside the birefringent coefficients $\kappa_D$ and $\kappa_z$.

We proceed using the same methodology and likelihood function described in Sec.~\ref{sec:methods}, and assume a merger rate described by Eq.~\eqref{eq:rate-model-sfr}.
Now, however, we do not fix the parameters $\{\alpha, \beta, z_p, R_0\}$ appearing in this equation, but elevate them as additional free parameters to be measured in our inference.
We also treat the maximum merger redshift $z_\mathrm{max}$ (previously set to $z_\mathrm{max}=15$) as an additional free parameter.
The direct detection of binary black hole mergers provides measurements of the local rate $R_0$ and the leading slope parameter $\alpha$, and excludes very small values of $z_\mathrm{p}$~\cite{Callister:2020arv,KAGRA:2021kbb,Callister:2023tgi}.
We adopt priors on these parameters, listed in Table~\ref{tab:priors}, consistent with these measurements.
As existing gravitational-wave measurements offer no information about $\beta$ or $z_\mathrm{max}$, we simply adopt uniform priors on these parameters over plausible ranges in order to marginalize over our systematic uncertainty.

\begin{table}[]
    \setlength{\tabcolsep}{8pt}
    \renewcommand{\arraystretch}{1.2}
    \centering
    \caption{Priors adopted on parameters governing the binary black hole merger rate, used when simultaneously inferring the merger rate alongside amplitude birefringence in Sec.~\ref{sec:variable-rate}.
    We use $N(\mu,\sigma)$ to denote a Gaussian prior with mean $\mu$ and standard deviation $\sigma$, and $U(a,b)$ to indicate a uniform prior on the interval $(a,b)$.
    }
    \begin{tabular}{r | l}
    \hline \hline
    Parameter & Prior\\
    \hline
    $\ln(R_0\cdot \mathrm{Gpc}^3\,\mathrm{yr})$ & $N(\ln 16,0.22)$ \\
    $\alpha$ & $N(3,1.5)$ \\
    $\beta$ & $U(1,10)$ \\
    $z_p$ & $U(0.5,4)$ \\
    $z_\mathrm{max}$ & $U(10,15)$ \\
    \hline
    \hline
    \end{tabular}
    \label{tab:priors}
\end{table}

When inferring only $\kappa_D$ and $\kappa_z$ under a fixed merger rate in Sec.~\ref{sec:fixed-rate}, we calculate posteriors simply iterating over a two-dimensional grid of possible birefringent coefficients.
When we now \textit{infer} the merger rate as well, the increased dimensionality of our target posterior makes this procedure infeasible.
Instead, we now sample from the joint posterior $p(\kappa_D,\kappa_z,R_0,\alpha,\beta,z_p,z_\mathrm{max})$ using the ``No U-Turn'' sampler~\cite{Hoffman2011} implemented in \textsc{numpyro}~\cite{numpyro1,numpyro2}, a probabilistic programming package built atop \textsc{jax}~\cite{jax}.

Our joint posterior on $\kappa_D$, $\kappa_z$, and parameters governing the black hole merger rate is shown in Fig.~\ref{fig:full_corner}.
We effectively recover our prior on the merger rate of black holes, with $R_0$ and $\alpha$ posteriors centered about their known values and uninformative posteriors on $z_p$, $\beta$, and $z_\mathrm{max}$.
The posteriors on these parameters, though, do not correlate strongly with posteriors on $\kappa_D$ and $\kappa_z$.
Our constraints on amplitude birefringence therefore  remain quantitatively similar to those found in Sec.~\ref{sec:fixed-rate}, despite our large uncertainty on the black hole merger rate.
The exact constraints on $\kappa_D$ and $\kappa_z$ are presented in Table~\ref{tab:kappa-constraints}.
Even after marginalization over the possible merger history of binary black holes, we find $|\kappa_D| \lesssim 0.2$ and $|\kappa_z|\lesssim0.3$ (each marginalized over the other parameter) or $|\kappa_D| \lesssim 0.1$ and $|\kappa_z|\lesssim0.2$ (fixing the other parameter to zero).

To illustrate the degree of uncertainty that these limits take into account, Fig.~\ref{fig:rate_constraints} shows the corresponding posterior on the binary black hole merger rate.
Each light trace corresponds to a single posterior sample drawn from Fig.~\ref{fig:full_corner}.
Similarly, Fig.~\ref{fig:OmgGW_constraints} shows the implied limits on the Stokes-I and Stokes-V of the stochastic gravitational-wave background, as in Fig.~\ref{fig:uniform_rate_omega}.

\section{Gravitational Wave constraints on Theory-Motivated Birefringence}
\label{sec:theory-motivated-results}
Having analyzed the birefringent scenario without any restriction, we now consider a scenario with theory-based priors imposed. With this method, we ensure that the birefringent corrections do not grow exponentially, but rather consider corrections which are leading order in deviations from GR and that map directly to parity-violating theories, following Eq.~\eqref{hRLmod}. The small parity violating contribution in this case does not allow us to place meaningful bounds with the current stochastic background non-observation, however we consider constraints that may be placed by a future detector network capable of observing a quieter background signal. 

\subsection{Theory Priors}
\label{sec:theorypriors}
As mentioned above, in this theory-motivated approach, we ensure that the linearization of the gravitational waves in $v_{\text{th}}(f)$ in Eq.~\eqref{eq:hrlLinear} remains valid for all parameter values sampled, and thus, any corrections remain as small deviations from GR at all times during the data analysis. This constraint will enter into the stochastic background parameter space as priors on the parameters $\alpha_{1}$ and $\beta_{1}$, 
\be 
\pi f \left(\frac{\alpha_{1}}{\Lambda_{\pv}}z  + \beta_{1} D_C\right) < 1. 
\ee 
For convenience, let us now rescale the parameter $\alpha_{1}$ such that 
\be 
\tilde{\alpha}_{1} = \frac{\alpha_{1}}{\Lambda_\pv}\frac{\hat{z}}{\hat{D}_c}, 
\ee 
where $\hat{z}$ and $\hat{D}_C$ are a reference redshift and comoving distance, respectively. We can then express our prior as 
\be 
\pi fD_C\left(\tilde{\alpha}_{1}\frac{\hat{D}_C}{D_C}\frac{z}{\hat{z}} + \beta_{1}\right) < 1.
\ee 
For convenience, we will take $\hat{D}_C = 1$ Gpc and $\hat{z} = 1$ and, as before, rescale the frequency to $f/100 \rm{Hz}$, such that we obtain 
\be 
\pi \frac{f}{100 \text{Hz}}\frac{D_C}{\text{Gpc}}\left(\tilde{\alpha}_{1}\frac{\text{Gpc}}{D_{C}}z + \beta_{1}\right) < 1.
\label{eq:priorparam}
\ee 
This is the prior that we will implement in our analysis below to ensure that the linear approximation of the waveform correction is valid at all times. To enforce this, while attempting to constrain the parameters $\alpha_{1}$ and $\beta_{1}$, we must consider the largest possible values for $f, D_C$, and $z$. We consider a maximum frequency of $f=10^3$ Hz. For the maximum redshift and comoving distance, we consider two cases: first the most conservative estimate for the background assuming a uniform merger rate, in which we take $z_{\max} \approx 6$ and ${D_C}_{\max} \approx 8.5\,\mathrm{Gpc}$, and then the more realistic delayed low-metallicity rate model, in which we take $z_{\max} \approx 10$ and ${D_C}_{\max} \approx 10\,\mathrm{Gpc}$.  With this in hand, the joint priors on $\alpha_{1}$ and $\beta_{1}$ are given by 
\begin{alignat}{2} 
&\left(\tilde{\alpha}_{1}+ 1.5\beta_{1}\right)_{\uni} &&\lesssim 0.005 \\
&\left(\tilde{\alpha}_{1}+ \beta_{1}\right)_\delayed &&\lesssim 0.003.
\label{eq:theoryprior}
\end{alignat}

We are considering an \textit{uninformative} prior on the parameters $\tilde{\alpha}_{1_0}$ and $\beta_{1_0}$ arising solely from EFT considerations, but one could alternatively consider a prior based on existing constraints  for some particular theories. As an example, consider Chern-Simons gravity, discussed in Sec.~\ref{sec:parity-violation}. Previous work has used binary pulsars to constrain CS gravity. Defining the parameter $\kappa_\cs^{-1} = (8\pi \alpha^\cs \dot{\vartheta})$ as in \cite{AliHaimoud:2011bk}, it was found that $\kappa_\cs^{-1} \lesssim 0.4$\,km. CS gravity maps to the $\alpha_{1_0}$ term in our constraint, which upon conversion to our conventions leads to
\be 
\tilde{\alpha}_{1} \lesssim 0.001. 
\ee 
Reaching a constraint of this level from the stochastic background with future detectors would thus place the strongest bounds to date on gravitational parity violation. Beyond CS gravity, additional work has studied amplitude birefringence with a correction proportional to our $\beta_{1}$ term, and placed a bound by stacking contraints from the  GWTC-3 catalogs \cite{Ng:2023jjt}. In this case, the parameterization used differs by a factor of $\pi$ from our $\beta_1$ term, and thus their result maps to the constraint  $\beta_{1} \lesssim 0.016$.
Note that the existing constraints on $\beta_{1}$ are indeed beyond the EFT bound in Eq.~\eqref{eq:theoryprior}. This is due to the fact that these constraints were found by studying events at much lower redshift than our maximum stochastic background binary distance. Thus, in considering only nearby events, the maximum value of the parity-violating parameters is allowed to be larger than what we obtain for the stochastic background, as long as the full constraint, Eq.~\eqref{eq:priorparam}, holds.

\subsection{Forecast of Constraints with Future Detectors}

\begin{figure*}
\includegraphics[width=0.5\textwidth]{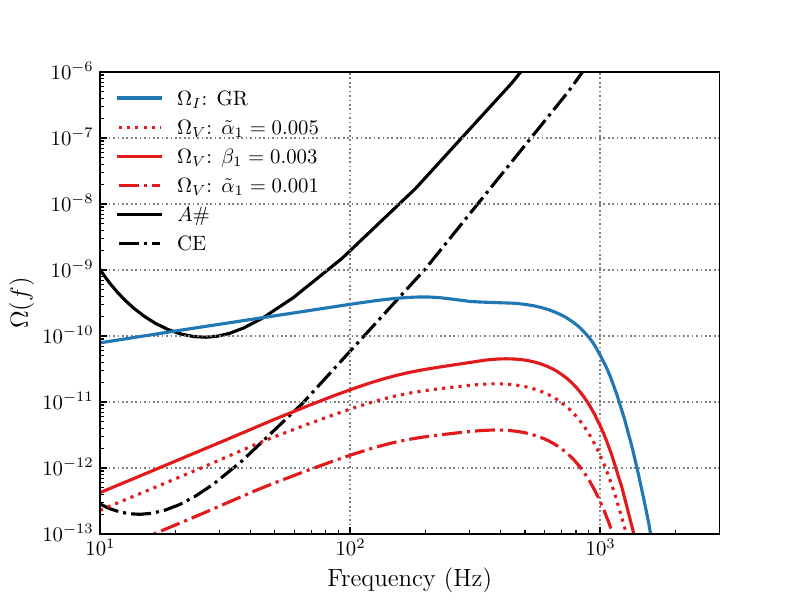}\hfill
\includegraphics[width=0.5\textwidth]{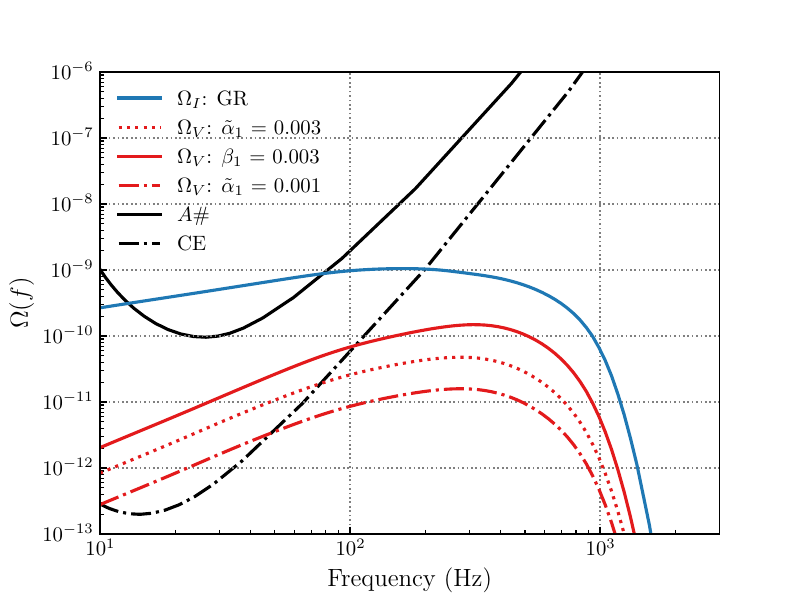}
    \caption{Forecasted sensitivity to the stochastic background, assuming a uniform merger rate (left) and delayed low-metallicity merger rate (right). We show explicitly the sensitivity curves for a two-detector baseline with $A\#$ and CE sensitivities, as well as several benchmark values for the birefringent parameters.}
    \label{fig:TheoryPlot}
\end{figure*}

 We now examine how well future detector networks with higher sensitivity may be able to tighten the bounds discussed in the previous section. We consider two concrete scenarios. Given that the projected LVK O4 sensitivity is unlikely to detect $\Omega_I(f)$, we first consider a two detector Hanford-Livingston configuration with projected ``A-sharp'' sensitivity (`A$^\sharp$'), as well as with Cosmic Exporer (`CE') sensitivity. We show the PI curves for each scenario in Fig.~\ref{fig:TheoryPlot}, compared to the expected GR value of $\Omega_I$, the maximum allowed theoretical value of $\Omega_V$ and the current strongest Chern-Simons constraint from binary pulsars. We show both the most pessimistic scenario of a uniform merger rate (left) and the more realistic scenario, considering $R_m^{\delayed}(z)$. 

We find that we are unlikely to place more restrictive bounds than \cite{Ng:2023jjt} until the sensitivity of third-generation detectors is obtained. In this case, the parity violating signal of $\Omega_V(f)$ will appear as a perturbation on top of $\Omega_I$, recalling that per Eq.~\eqref{eq:OmegaIth} there is no parity violating correction at linear order to the expected $\Omega_I$ in GR. Then, to extract a constraint on birefringence, one will need to precisely measure the expected GR signal to determine if there is any excess. Our projections predict that such a constraint on the  birefringent parameters $\alpha_1$ and $\beta_1$ will be competitive. Even in the more pessimistic scenario for the merger rate shown in the left panel of Fig.~\ref{fig:TheoryPlot}, the sensitivity reached with a CE detector network is nearly enough to match the constraint on Chern-Simons gravity from binary pulsars (corresponding to $\tilde{\alpha}_1 = 0.001$). Taking a more optimistic view of the merger rate, as shown in the right panel of Fig.~\ref{fig:TheoryPlot}, the overall amplitude of the background is increased, and thus, we will be able to set better constraints. In this case, the constraints on the parameters $\tilde{\alpha}_1$ and $\beta_1$ will be $\mathcal{O}(10^{-4})$ from a non-detection of $\Omega_V$ with a CE network. This result is an order of magnitude stronger than the binary pulsar result, thus would be the most stringent constraint on gravitational parity violation to date.

\section{Discussion and Conclusions}
\label{sec:discussion}

We have shown that the stochastic background provides a new test of parity violation.
We have used the non-detection of the stochastic gravitational wave background from compact binaries through LVK observations to place a constraint on gravitational parity violation.
From a phenomenological perspective, we have placed a new independent constraint on birefringence parameters:  $(\kappa_D, \kappa_V)  \lesssim \mathcal{O}(10^{-1})$. This is competitive with current constraints from individual sources, and we expect it to improve with observations from LVK O4. We have furthermore forecasted the power of future detectors to constrain gravitational parity violation with stochastic background observations when theory-motivated priors are imposed. We found that third-generation detectors will be able to improve the constraint by $\sim$ two orders of magnitude, yielding new stringent bounds on parity-violating theories.

Additionally, the stochastic background has the potential to allow us to probe parity violation beyond compact binaries. Supermassive black hole mergers, which are thought to be responsible for the stochastic background signal found by pulsar timing arrays, should also lead to birefringent gravitational waves if gravity is indeed parity violating. Although currently the PTA observations are not sensitive to any parity or polarization in the signal, it is possible that in the future such determinations will be possible, as suggested by~\cite{Kato:2015bye,Belgacem:2020nda, Sato-Polito:2021efu}. It has also been suggested that astrometry can be used as a complementary technique to search for parity violation in the nanohertz regime \cite{Liang:2023pbj}. Furthermore, the expected \textit{cosmological} stochastic background arising from primordial processes may also lead to distinctive parity-violating signatures. For example, it has been suggested that modified gravity theories, such as Chern-Simons gravity, could lead to a parity-violating galaxy four-point function from processes during inflation \cite{Creque-Sarbinowski:2023wmb}. This type of interaction, as well as other early universe processes, could additionally lead to gravitational waves that undergo birefringence in their propagation, which can then be probed with the stochastic background. We leave these prospects and others to future work.

\section*{Acknowledgements}
We thank Wayne Hu and Jacob Golomb for useful comments. 
T.~C.~ is supported by the Eric and Wendy Schmidt AI in Science Postdoctoral Fellowship, a Schmidt Futures program. The work of L.~J.~ and D.~E.~H~ is supported by the Kavli Institute for Cosmological Physics at the University of Chicago through an endowment from the Kavli Foundation and its founder Fred Kavli. D.~E.~H~ is further supported by NSF grants AST-2006645 and PHY-2110507. N.~Y.~is supported from the Simons Foundation through Award No. 896696 and the National Science Foundation (NSF) Grant No. PHY-2207650. This material is based upon work supported by NSF's LIGO Laboratory which is a major facility fully funded by the National Science Foundation.\\

\noindent \textit{\bf Data and code availability}:
The code used for this study is hosted on GitHub at \url{https://github.com/tcallister/stochastic-birefringence}, and data produced by our analyses can be download from Zenodo at
 \url{https://zenodo.org/doi/10.5281/zenodo.10384998}.
\\

\noindent \textit{\bf Software}:
{\tt arviz}~\cite{arviz_2019}, {\tt astropy}~\cite{astropy1,astropy2}, {\tt h5py}~\cite{h5py}, {\tt jax}~\cite{jax}, {\tt matplotlib}~\cite{Hunter:2007}, {\tt numpy}~\cite{numpy}, {\tt numpyro}~\cite{numpyro1,numpyro2}, {\tt scipy}~\cite{scipy}, {\tt stochastic.m}~\cite{stochastic-dot-m}.

\appendix 

\section{Investigating the ``spike''}
\label{app:spike}

\begin{figure*}
    \includegraphics[width=0.95\textwidth]{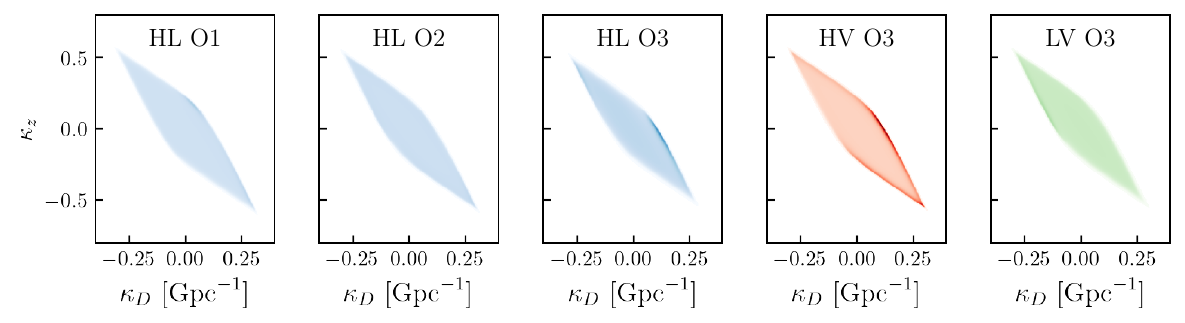}
    \caption{
    Posteriors on $\kappa_D$ and $\kappa_z$ obtained using individual baseline pairs from individual observing runs.
    We specifically show results from the Hanford-Livingston (HL) baseline during the the first, second, and third observing runs (O1, O2, and O3, respectively), and the Hanford-Virgo (HV) and Livingston-Virgo (LV) baseline during O3.
    We adopt $R_m^{\rm delayed}(z)$ from Eq.~\eqref{eq:rate-model-delayed} as our model for the binary merger rate, as in the right corner plot of Fig.~\ref{fig:SFR_results}.
    We see signs of the elevated edge or ``spike'' in both the Hanford-Livingston and Hanford-Virgo data during O3.
    We do not see any features arising in O1 or O2, nor in the Livingston-Virgo O3 data.
    }
    \label{fig:baseline-comparison}
\end{figure*}

\begin{figure*}
    \includegraphics[width=0.95\textwidth]{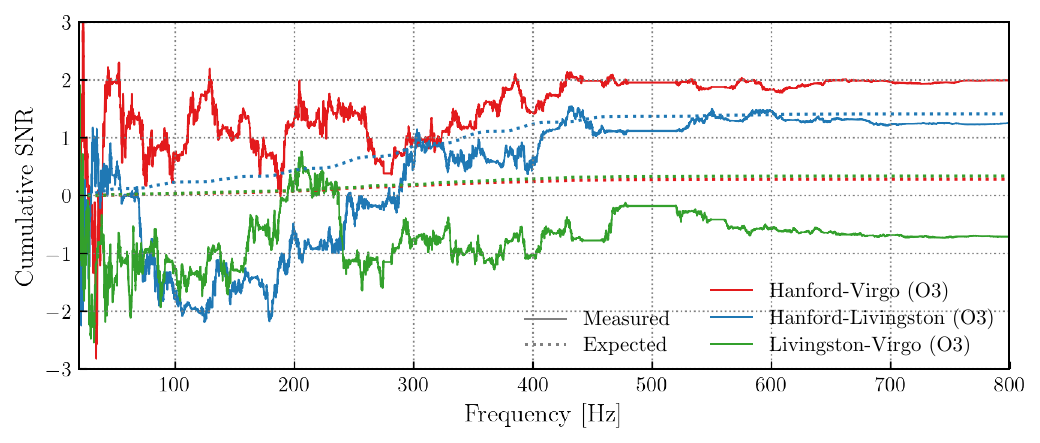}
    \caption{
    Cumulative signal-to-noise ratios (SNRs) of the Hanford-Livginston, Hanford-Virgo, and Livingston-Virgo baselines during O3.
    Dotted curves show the anticipated increase in SNR as we integrate across frequencies, assuming the presence of a birefringent stochastic background with $\kappa_D=\MaxLikelihoodKappaDC$ and $\kappa_z=\MaxLikelihoodKappaZ$, the maximum posterior values from the right-hand side of Fig.~\ref{fig:SFR_results}.
    The solid lines, meanwhile, show the \textit{actual} measured SNRs.
    There are no clear narrowband features that would explain the elevated ``ridge'' features seen in Fig.~\ref{fig:baseline-comparison} above.
    Instead, all three baselines appear to accumulate SNR as a random walk, consistent with expected behavior in Gaussian noise.
    At the same time, there is an apparent mismatch between the \textit{magnitudes} of predicted and observed SNRs.
    According to our signal model (dotted lines), the Hanford-Livingston baseline is expected to witness the loudest signal, but it is instead the Hanford-Virgo baseline that contains the largest SNR.
    See Fig.~\ref{fig:target-snrs} for further details.}
    \label{fig:cumulative-snrs}
\end{figure*}

\begin{figure}
    \includegraphics[width=0.45\textwidth]{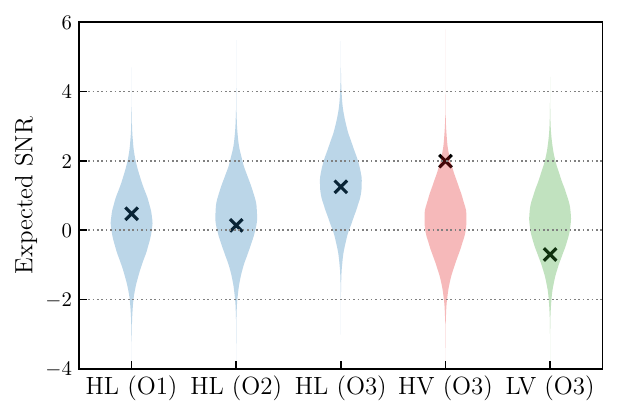}
    \caption{Measured signal-to-noise ratios (SNRs) from each detector baseline and observing run (black crosses), compared with the expected probability distributions of SNRs under the best-fitting signal model (violins).
    Expected SNRs are computed by adopting $\kappa_D=\MaxLikelihoodKappaDC$ and $\kappa_z=\MaxLikelihoodKappaZ$, the maximum posterior values obtained when assuming a black hole merger history that traces delayed low-metallicity star formation (right-hand side of Fig.~\ref{fig:SFR_results}).
    Each probability distribution is then obtained by considering possible realizations of random, Gaussian noise.
    Data from the Hanford-Livingston baseline yields SNRs consistent with expectation.
    The Hanford-Virgo baseline, however, appears to contain anomalously high SNR.
    As discussed in the text, however, the high SNR of Hanford-Virgo is not statistically significant.
    If a birefringent stochastic background were indeed present in the data, random noise fluctuations produce an SNR as loud as Hanford-Virgo's with probability $P=\ProbAboveWithTrials$.
    And in the complete absence of a stochastic signal, noise fluctuations will still produce an equally loud SNR with $P=\ProbAboveWithTrialsNoise$.
    Thus the measured SNRs remain consistent with a non-detection of a stochastic background.
    }
    \label{fig:target-snrs}
\end{figure}

Although we do not constrain $\kappa_D$ or $\kappa_z$ away from zero, we do nevertheless see an unexpected feature in our posteriors: a sharp, elevated edge in the two-dimensional $\kappa_D$-$\kappa_z$ posterior.
When conditioned on $\kappa_D = 0$ and/or $\kappa_z = 0$, this feature translates into a sharply elevated spike in the marginal posteriors on $\kappa_D$ and $\kappa_z$.
In this appendix, we explore the nature of this feature and its consistency (or inconsistency) with a signal of astrophysical origin.

To begin, Fig.~\ref{fig:baseline-comparison} shows joint posteriors on $\kappa_D$ and $\kappa_z$ obtained by analyzing only one baseline and one observing run at a time.
We assume that the binary black hole merger rate follows the delayed low-metallicity star formation rate (Eq.~\eqref{eq:rate-model-delayed}), as in the right corner plot of Fig.~\ref{fig:SFR_results}.
We see signs of the elevated spike in two of the individual datasets.
First, there is a marginal feature near $\kappa_D\approx 0.1$ arising from Hanford-Livingston cross-correlation data during O3 (HL O3).
Second, and far more prominently, a very pronounced feature appears in the Hanford-Virgo O3 data (HV O3), tracing the full right-hand edge of the posterior.

These observations together raise several questions:
Does the ``spike'' originate from particular frequency intervals with anomalous or elevated cross-power, or is it broadband in nature?
Furthermore, given known instrumental sensitivities and baseline geometries, should we expect a real stochastic background to behave as in Fig.~\ref{fig:baseline-comparison}, that is, with no discernible signal in Livingston-Virgo, a slight feature in Hanford-Livingston, and a comparably loud excess in Hanford-Virgo?

As a first step, we can explore the frequency dependence of the tentative signal.
Equation~\eqref{eq:inner-product} in Sec.~\ref{sec:methods} defined an inner product, $(A|B)$, in terms of which the likelihood was expressed.
The same inner product gives the signal-to-noise ratio (SNR) of a stochastic background measurement.
Let $\hat C(f)$ be a cross-correlation measurement from a particular baseline pair~\footnote{We previously used the notation $\hat C_{ij}(f)$ with $i$ and $j$ labeling detectors, but here suppress these subscripts for simplicity.} with overlap reduction functions $\gamma^A(f)$, where $A\in\{I,V\}$.
Also, let $\Omega_A^M(f)$ be some \textit{model template} for the stochastic energy-density $\Omega_A(f)$ in each polarization $A$.
Then, the measured SNR associated with this model is
    \be
    \mathrm{SNR} = \frac{(\hat C|\gamma^A\Omega_A^M)}{\sqrt{(\gamma^{A'} \Omega_{A'}^M|\gamma^{A''} \Omega_{A''}^M)}}.
    \ee
We have again made use of an Einstein-summation-like convention in which $\gamma^A(f) \Omega_A^M(f) = \gamma^I(f) \Omega_I^M(f) + \gamma^V(f) \Omega_V^M(f)$.
Recall that the expectation value of our cross-correlation statistic is $\langle \hat C(f) \rangle = \gamma^A(f) \Omega_A(f)$.
If our model for the stochastic background  is actually correct, with $\Omega_A(f) = \Omega_A^M(f)$, then the expectation value of our signal-to-noise ratio is
    \be
    \left\langle \mathrm{SNR} \right\rangle
        = \sqrt{(\gamma^{A} \Omega_{A}^M|\gamma^{A'} \Omega_{A'}^M)}.
    \ee

Having defined the total signal-to-noise ratio of a signal, we can now investigate how this signal-to-noise \textit{accumulates} as we integrate across frequencies.
A true astrophysical signal should be broadband in nature, yielding a signal-to-noise that accumulates gradually across a range of frequencies.
In contrast, terrestrial contamination should likely (although not necessarily) manifest as a \textit{narrowband} signal, in which the measured signal-to-noise is dominated by a small collection of frequency bins.

With this in mind, let us define the \textit{cumulative inner product}
\be
(A|B)|_{f} = \int_0^f df' \frac{A(f') B^*(f') + A^*(f') B(f')}{\sigma^2_{ij}(f')},
\label{eq:inner-product-fmax}
\ee
which is identical in form to Eq.~\eqref{eq:inner-product} but now includes only frequencies below $f$.
Then, the \textit{cumulative SNR}, obtained if we had access only to data below $f$, is
    \be
    \mathrm{SNR}(f) = \frac{(\hat C|\gamma^A\Omega_A^M)|_{f}}{\sqrt{(\gamma^{A'} \Omega_{A'}^M|\gamma^{A''} \Omega_{A''}^M)|_{f}}},
    \label{eq:snr-f}
    \ee
with expectation value
    \be
    \left\langle \mathrm{SNR}(f) \right\rangle
        = \sqrt{(\gamma^{A} \Omega_{A}^M|\gamma^{A'} \Omega_{A'}^M)|_{f}}.
    \label{eq:snr-f-expected}
    \ee

When assuming that black hole mergers trace the delayed, low-metallicity star formation rate (right corner plot in Fig.~\ref{fig:SFR_results}), the resulting posterior is maximized at $\kappa_D=\MaxLikelihoodKappaDC$ and $\kappa_z=\MaxLikelihoodKappaZ$.
Figure~\ref{fig:cumulative-snrs} shows the cumulative SNRs for the Hanford-Livingston, Hanford-Virgo, and Livingston-Virgo baselines during O3 using these maximum posterior values.
Solid curves show the actual recovered SNR and dotted curves give the expectation values for each baseline (Eqs.~\eqref{eq:snr-f} and ~\eqref{eq:snr-f-expected}, respectively).

The Hanford-Virgo baseline contains the loudest signal, with $\mathrm{SNR}\approx \HVSNR$ after integrating across all frequencies.
This is consistent with the fact that, in Fig.~\ref{fig:baseline-comparison}, it is also the Hanford-Virgo baseline that exhibits the most structure in its posterior.
The Hanford-Livingston baseline gives a weaker $\mathrm{SNR}\approx \HLSNR$, and the Livingston-Virgo baseline contains a statistically insignificant $\mathrm{SNR}=\LVSNR$.
In all three cases, there are no clear narrowband features dominating the total signal-to-noise ratio.
The Hanford-Virgo baseline, in particular, appears to accumulate SNR as a random walk, as expected in well-behaved Gaussian data.
Note that the distinctive ``flat'' stretches mark frequencies in which all data has been masked due to known instrumental artifacts~\cite{LSC:2018vzm,KAGRA:2021kbb} and so contribute zero signal.

Although the cumulative SNRs appear well-behaved, are the total SNRs measured in each baseline consistent with expectation?
The dashed lines in Fig.~\ref{fig:cumulative-snrs} indicate that, under the chosen parameters, we expect the Hanford-Livingston baseline to witness the loudest signal, with near-zero SNR in the Hanford-Virgo and Livingston-Virgo baselines.
An alternative illustration is given in Fig.~\ref{fig:target-snrs}, which shows the probability distributions of SNRs across each baseline and observing run (violins) and the actual SNR measured in each case (black crosses), again assuming $\kappa_D=\MaxLikelihoodKappaDC$ and $\kappa_z=\MaxLikelihoodKappaZ$.
The data do not appear to match expectations.
Although the SNR of Hanford-Livinston O3 data lies in the anticipated range, the SNR from the Hanford-Virgo O3 baseline is much larger than predicted under the best-fitting signal model.
And at the same time, the Livingston-Virgo O3 baseline yields somewhat lower SNR than expected.

Since the SNRs of stochastic cross-correlation measurements are well-described by Gaussian statistics, we can compute how likely this situation is to arise by chance.
In the presence of a birefringent stochastic background with $\kappa_D=\MaxLikelihoodKappaDC$ and $\kappa_z=\MaxLikelihoodKappaZ$, the probability that noise fluctuations produce a signal as loud as that seen in Hanford-Virgo during O3 is $P=\ProbHVSNR$. 
Moreover, the probability that noise fluctuations would simultaneously give SNRs greater than in Hanford-Virgo and smaller than in Livingston-Virgo during O3 is $P=\ProbHVLVSNR$.
Both significance estimates, though, neglect trials factors accounting for the fact that, with five distinct datasets, we increase the likelihood that at least \textit{one} of them individually exhibits a statistically unlikely outcome.
With five datasets, the probability that at least one exhibits an SNR fluctuation as large as that in Hanford-Virgo is $P=\ProbAboveWithTrials$. 
And the probability that at least one exhibits an upward SNR fluctuation consistent with Hanford-Virgo's \textit{and} at least another has SNR fluctuating downward consistent with Livingston-Virgo is $P=\ProbAboveAndBelowWithTrials$.

Both the above probabilities indicate that the measured SNRs are statistically consistent with random fluctuations atop a birefringent stochastic background.
Hence, we cannot \textit{rule out} a potential signal on the basis of SNR consistency.
Along these lines, though, do the observed SNRs even require a signal at all, or are they also consistent with pure Gaussian noise?
In the complete absence of an astrophysical signal, the probability that Gaussian noise fluctuations produce an SNR as large as Hanford-Virgo's during O3 in at least one of five baselines is $P=\ProbAboveWithTrialsNoise$.
And the probability that noise produces SNRs both above and below those in Hanford-Virgo and Livingston-Virgo, respectively, is $P=\ProbAboveAndBelowWithTrialsNoise$.
The measured SNRs therefore remain statistically consistent with random instrumental noise.

\section{Mass distribution}
\label{app:MassDist}

\begin{figure}
    \includegraphics[width=0.48\textwidth]{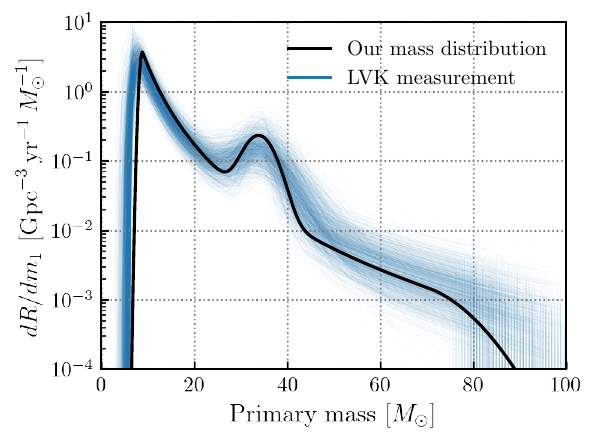}
    \caption{
    The differential black hole merger rate as a function primary mass that we adopt in this work (black).
    Specifically, we show the differential merger rate at $z=0$; throughout the text we explore several different models for how this rate subsequently evolves with redshift.
    The black hole mass function is described by Eq.~\eqref{eq:simple-mass-distribution} with parameters given in  Eq.~\eqref{eq:mass-params}.
    For comparison, we show in blue the posterior obtained on the black hole mass spectrum in Ref.~\cite{KAGRA:2021duu} using their \textsc{Power Law \& Peak} mass model; each trace corresponds to one individual draw from their posterior.
    Note that, within Ref.~\cite{KAGRA:2021duu}, the differential merger rate is shown at a reference redshift of $z=0.2$, rather than $z=0$ as shown here, yielding merger rates that appear slightly elevated relative to those shown here.
    }
    \label{fig:mass-comparison}
\end{figure}

In our analyses, we assume black hole masses to follow a distribution consistent with hierarchical measurements of the black hole mass function~\cite{KAGRA:2021duu}.
In this Appendix, we provide a detailed description of the precise mass distribution used.

We take primary masses to follow a mixture between a power law and a Gaussian peak:
    \be
    \begin{aligned}
    \phi(m_1) = &(1-f_p)  \frac{1+\lambda}{(100\,M_\odot)^{1+\lambda} - (2\,M_\odot)^{1+\lambda}} m_1^{\lambda} \\
        & \hspace{1cm} + \frac{f_p}{\sqrt{2\pi\sigma^2}}
            \mathrm{exp}\left[-\frac{(m_1-\mu)^2}{2 \sigma^2}\right].
    \end{aligned}
    \label{eq:simple-mass-distribution}
    \ee
Here, $\mu$ and $\sigma^2$ give the mean and variance of the Gaussian peak, $\lambda$ is the slope of the power-law component, and $f_p$ governs the mixture fraction between each component.
The power-law is normalized over the interval $2\,M_\odot \leq m_1 \leq 100\,M_\odot$.
Additionally, we assume that the primary mass distribution is smoothly truncated to zero beyond a minimum mass $m_\mathrm{min}$ and maximum mass $m_\mathrm{max}$, such that the full mass distribution is of the form
    \begin{equation}
    p(m_1) \propto
        \begin{cases}
        \phi(m_1)\,\mathrm{Exp}\left[-\frac{(m_1-m_\mathrm{min})^2}{2\delta m_\mathrm{min}^2}\right] & \left(m_1\leq m_\mathrm{min}\right) \\
        \phi(m_1) & \left(m_\mathrm{min}< m_1 \leq m_\mathrm{max}\right) \\
        \phi(m_1)\,\mathrm{Exp}\left[-\frac{(m_1-m_\mathrm{max})^2}{2\delta m_\mathrm{max}^2}\right] & \left(m_\mathrm{max}< m_1\right),
        \end{cases}
    \end{equation}
where the parameters $\delta m_\mathrm{min}$ and $\delta m_\mathrm{max}$ control the scale over which the truncations occur.
We assume that secondary masses are distributed as a power law between $2\,M_\odot$ and $m_1$,
    \be
    p(m_2|m_1) = \frac{1+\beta_q}{m_1^{1+\beta_q} - (2\,M_\odot)^{1+\beta_q}} m_1^{\beta_q},
    \ee
with slope $\beta_q$.

We fix the parameters of these distributions to values consistent with the black hole mass distribution reported in Ref.~\cite{KAGRA:2021duu}:
    \be
    \begin{aligned}
    m_\mathrm{max} &= 70\,M_\odot \\
    m_\mathrm{min} &= 9\,M_\odot \\
    \delta m_\mathrm{max} &= 10\,M_\odot \\
    \delta m_\mathrm{min} &= 0.5\,M_\odot \\
    \mu &= 34\,M_\odot \\
    \sigma &= 3\,M_\odot \\
    \lambda &= -3.8 \\
    f_p &= 10^{-2.7} \\
    \beta_q &= 2\,.
    \end{aligned}
    \label{eq:mass-params}
    \ee
Figure~\ref{fig:mass-comparison} shows the differential merger rate given by the above parameters (black), as well as the posterior on the black hole mass function from Ref.~\cite{KAGRA:2021duu}.
While our model for the primary mass distribution is not strictly the same as the \textsc{Power Law \& Peak} model adopted in~\cite{KAGRA:2021duu} (which adopts a slightly different function form for high- and low-mass truncations), our chosen model agrees well with their measurements. 

\section{Comment on Signal Normalization Conventions}

\begin{figure*}
    \includegraphics[width=0.95\textwidth]{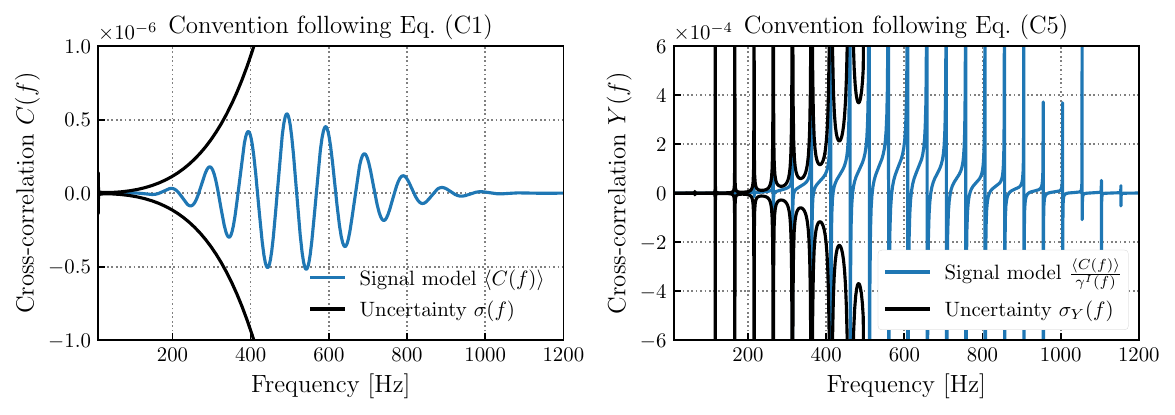}
    \caption{
    Illustration of two different signal normalization conventions used to model and search for the stochastic gravitational-wave background.
    \textit{Left}: The convention used in this paper, following Eq.~\eqref{eq:cross-corr-statistic-again}.
    Under this convention, the cross-correlation statistic is normalized such that its expectation value is $\gamma^I(f) \Omega_I(f) + \gamma^V(f)\Omega_V(f)$.
    The blue curve shows an example signal model, normalized in this fashion.
    Under convention~\eqref{eq:cross-corr-statistic-again}, meanwhile, the cross-correlation statistic has variance $\sigma^2(f)$ that is a smooth function of frequency.
    The black curves show $\pm \sigma(f)$ under this convention.
    \textit{Right}:
    A more common convention, following Eq.~\eqref{eq:cross-corr-statistic-alt}, in which the cross-correlation statistic is divided by the Stokes-I overlap reduction function, yielding an alternative statistic $\hat Y(f) = \hat C(f)/\gamma^I(f)$.
    Because $\gamma^I(f)$ passes through zero, this yields a signal model that periodically diverges to infinity.
    The corresponding variance under this convention is given by $\sigma_Y(f) = \sigma(f)/(\gamma^I(f))^2$.
    This too diverges to infinity at the roots of the overlap reduction function.
    Data products released by the LIGO-Virgo-KAGRA collaboration follow this second convention.
    }
    \label{fig:signal_conventions}
\end{figure*}

In this final appendix, we highlight a pitfall we encountered over the course of this study, flagging it as a point of caution for other researchers analyzing publicly released cross-correlation measurements of the stochastic gravitational-wave background.

In our study, we defined the cross-correlation statistic~\footnote{We previously called this quantity $\hat C_{ij}(f)$, but in this appendix will neglect the subscripts $i$ and $j$.} 
\be
\hat C(f) = \frac{1}{T} \frac{20 \pi^2}{3 H_0^2} f^3 \tilde s_i(f) \tilde s_j^*(f)
\label{eq:cross-corr-statistic-again}
\ee
(Eq.~\eqref{eq:cross-corr-statistic} in the main text)
between data $\tilde s_i$ and $\tilde s_j$ from detectors $i$ and $j$.
In the presence of a polarized gravitational-wave background, the expectation value of this statistic is (again repeating equations for convenience)
\be
\langle \hat C(f) \rangle = \gamma^I(f) \Omega_I(f) + \gamma^V(f) \Omega_V(f)
\label{eq:sensible-convention-signal}
\ee
and its variance is
\be
\langle \hat C(f) \hat C(f') \rangle = \delta(f-f') \sigma^2(f),
\ee
where
\be
\sigma^2(f) = \frac{1}{T} \left(\frac{10\pi^2}{3 H_0^2}\right)^2 f^6 P_i(f) P_j(f).
\label{eq:sensible-convention-noise}
\ee

Equation~\eqref{eq:cross-corr-statistic-again} is not the standard convention.
More common is the alternate convention 
    \be
    \hat Y(f) = \frac{1}{T} \frac{20 \pi^2}{3 H_0^2} \frac{f^3 \tilde s_i(f) \tilde s_j^*(f)}{\gamma^I(f)}
    \label{eq:cross-corr-statistic-alt}
    \ee
in which we have divided by the Stokes-I overlap reduction function $\gamma^I(f)$ for the given baseline.
In this convention, the expectation value of the cross-correlation statistic is
    \be
    \langle \hat Y(f) \rangle = \Omega_I(f) + \frac{\gamma^V(f)}{\gamma^I(f)} \Omega_V(f)
    \ee
and its variance is 
    \be
    \langle \hat Y(f) \hat Y(f') \rangle = \delta(f-f') \sigma_Y^2(f),
    \ee
where
    \be
    \sigma^2_Y(f) = \frac{1}{T} \left(\frac{10\pi^2}{3 H_0^2}\right)^2 \frac{f^6 P_i(f) P_j(f)}{\left(\gamma_I(f)\right)^2}.
    \ee
Equation~\eqref{eq:cross-corr-statistic-alt} is motivated by the fact that most analyses operate under the assumption that the stochastic background is unpolarized.
In this case, $\Omega_V(f)$ is presumed to be zero and $ \langle \hat Y(f) \rangle = \Omega_I(f)$, such that $\hat Y(f)$ is a direct estimator of the total energy-density of the stochastic background.

\begin{figure*}
    \includegraphics[width=0.95\textwidth]{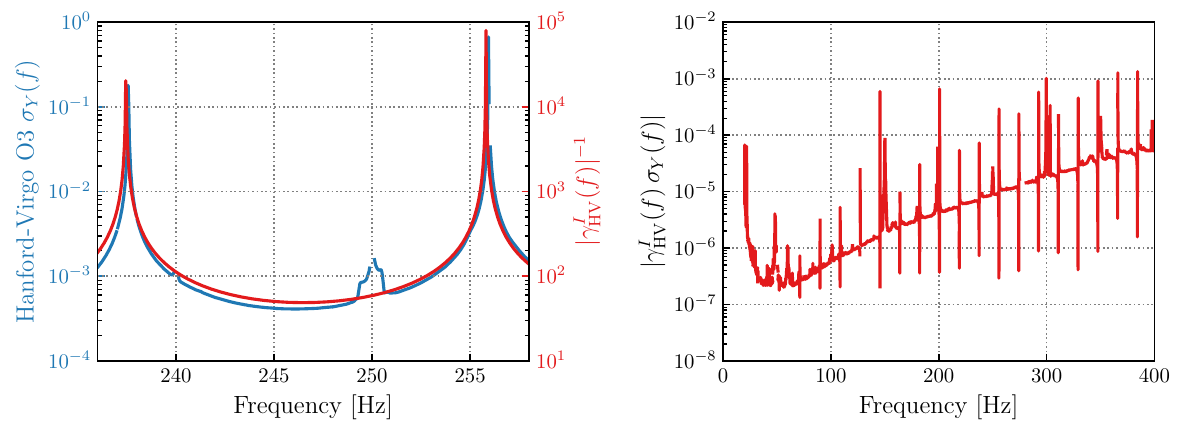}
    \caption{
    Illustration of numerical issues that can arise when analyzing public data normalized following convention~\eqref{eq:cross-corr-statistic-alt}.
    \textit{Left}: Shown in blue are is the uncertainty spectrum $\sigma_Y(f) = \sigma(f)/\gamma^I(f)$ for the Hanford-Virgo baseline during O3, as released by the LIGO-Virgo-KAGRA collaboration.
    Because this follows the convention in which the cross-correlation statistic has been divided by $\gamma^I(f)$, analysts who define and test their own gravitational-wave background models must correspondingly multiply their models by $1/\gamma^I(f)$.
    Our \texttt{python}-based calculation of this factor is shown in red.
    Both $\sigma_Y(f)$ and $1/\gamma^I(f)$ diverge to infinity at the roots of the overlap reduction function.
    At issue is the fact that different calculations of the overlap reduction function place roots in slightly different locations.
    Our \texttt{python}-based calculation, for example, is seen to yield infinite divergences at fractionally different locations than the \texttt{matlab} calculation used to produce $\sigma_Y(f)$.
    \textit{Right}: The ratio between the blue and red curves on the left.
    When following convention~\eqref{eq:cross-corr-statistic-alt}, it is this ratio that enters into likelihood and SNR calculations.
    The failure to produce divergences in exactly the same locations with the same slopes produces in the left-hand subplot yields the sharp comb of features seen in the right-hand side, features that can non-negligibly perturb the total recovered SNR and/or likelihood, as illustrated below.
    }
    \label{fig:orf-cancellation-bad}
\end{figure*}

\begin{figure*}
    \includegraphics[width=0.95\textwidth]{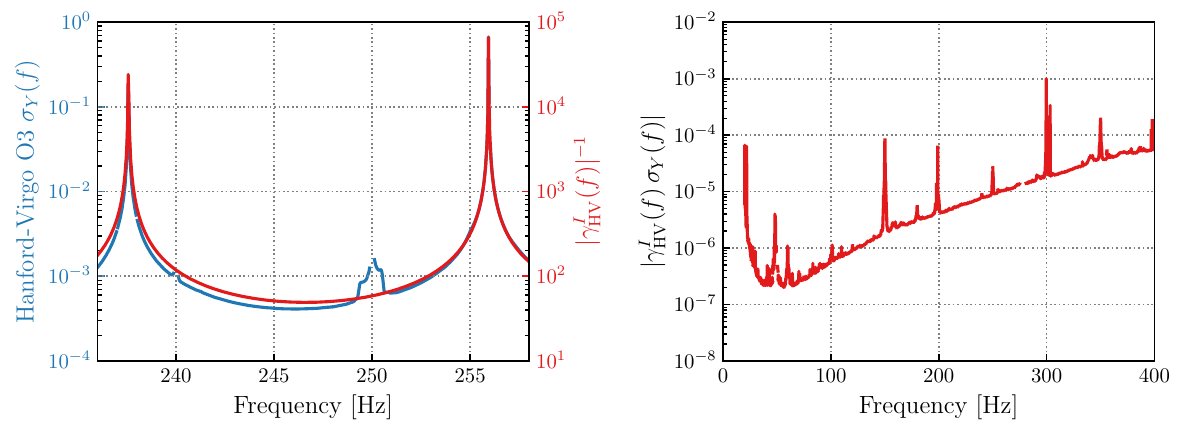}
    \caption{
    As in Fig.~\ref{fig:orf-cancellation-bad}, but showing the correct intended behavior.
    \textit{Left}: The same as the left-hand panel of Fig.~\ref{fig:orf-cancellation-bad}, but now showing the factor $1/\gamma^I(f)$ as computed using exactly the same \texttt{matlab} routine used internally to create $\sigma_Y(f)$.
    \textit{Right}: Exact agreement between the divergences in $\sigma_Y(f)$ and $1/\gamma^I(f)$ now produces a smooth ratio between the two curves, enabling well-behaved likelihood and SNR calculations.
    Only by using identical algorithms with identical software packages were we able to obtain this well-behaved result.
    Otherwise, failure to imperfectly undo the ``division by zero'' in the definition of $\sigma_Y(f)$ produced spurious SNR estimates, as shown in Fig.~\ref{fig:cumulative_snrs_orf_comparison} below.
    This danger is mitigated if data are produced entirely following convention~\eqref{eq:cross-corr-statistic-again}, in which signal and noise estimates are never divided by $\gamma_I(f)$ and hence never contain infinite divergences that must be exactly undone by downstream analysts.
    }
    \label{fig:orf-cancellation-good}
\end{figure*}

\begin{figure*}
    \includegraphics[width=0.95\textwidth]{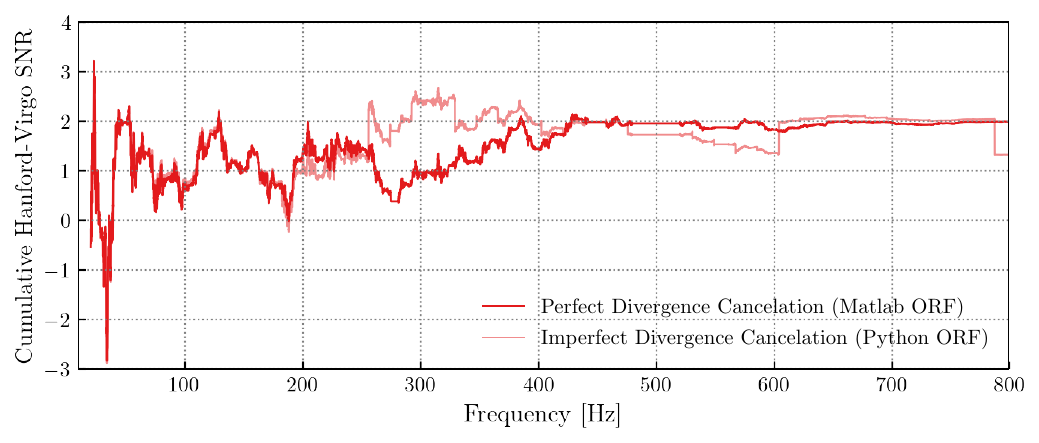}
    \caption{
    A comparison of the cumulative signal-to-noise ratios estimated using imperfect (Fig.~\ref{fig:orf-cancellation-bad}) and perfect (Fig.~\ref{fig:orf-cancellation-good}) cancellation of the divergences in $\sigma_Y(f)$.
    Imperfect cancellation occurs when naively using overlap reduction functions calculated in \texttt{python}; perfect cancellation requires instead using overlap reduction functions calculated in \texttt{matlab} using the exact same routine used to compute $\sigma_Y(f)$.
    In the case of imperfect cancellation, the resulting combs seen in $\gamma^I(f)\sigma_Y(f)$ (specifically the \textit{downward} spikes extending to zero) result in a small number frequency bins that are spuriously and significantly upweighted.
    As we integrate across frequencies, the cumulative SNR exhibits random and discontinuous jumps as we encounter these upweighted frequencies, such as those that can be seen at approximately $250$ and $330$~Hz.
    In the example shown, these random jumps happen to nearly cancel one another, such that the total SNR is comparable in both cases.
    This cancelation is not guaranteed, however, and in alternative cases could perturb SNR estimates by significant amounts.
    At $f\approx 300$, for example, the two SNR estimates differ significantly.
    Differences of this magnitude could lead to false claims of detection, or false dismissal of an actual stochastic gravitational-wave background.
    }
    \label{fig:cumulative_snrs_orf_comparison}
\end{figure*}

These two conventions are sketched in Fig.~\ref{fig:signal_conventions}.
The left-hand side illustrates an example signal model (blue) and the associated uncertainty spectrum (black) defined in accordance with Eq.~\eqref{eq:cross-corr-statistic-again}.
The right-hand side corresponds to the same data, but now normalized following Eq.~\eqref{eq:cross-corr-statistic-alt}, having been divided by $\gamma^I(f)$.
Because the overlap reduction function $\gamma^I(f)$ is oscillatory about zero, when we divide by $\gamma^I(f)$, we correspondingly introduce infinite divergences in both the signal model $\langle Y(f) \rangle = \langle C(f)/\gamma^I(f)\rangle$ and the uncertainty $\sigma_Y(f) = \sigma(f)/\gamma^I(f)$.
These are easily seen in the right-hand side of Fig.~\ref{fig:signal_conventions}.

In principle, it should not matter which convention is adopted, provided we then proceed self-consistently with this convention.
Likelihood and SNR calculations do not depend directly on cross-correlation measurements, model spectra, and uncertainty spectra, but only on the ratios $\hat C(f)/\sigma(f) = \hat Y(f)/\sigma_Y(f)$ and $\langle C(f)\rangle/\sigma(f) = \langle Y(f)\rangle/\sigma_Y(f)$, which are nominally identical under either convention.
The fact that our signal model diverges to infinity under convention~\eqref{eq:cross-corr-statistic-alt} is therefore fine, provided that $\sigma_Y(f)$ diverges to infinity at precisely the same locations and with the same speed.

In practice, however, the fact that data is normalized according to Eq.~\eqref{eq:cross-corr-statistic-alt} gives rise to scenarios in which imperfect numerical convergence spuriously and non-negligibly impacts results.
The issue is that publicly-available cross-correlation data provide $\sigma_Y(f)$ spectra that already include division by $\gamma^I(f)$, while users must \textit{independently} recalculate and divide their signal model by $\gamma^I(f)$ to ensure consistent normalization.
\textit{If the user-calculated $\gamma^I(f)$ differs, even fractionally, from the overlap reduction function used to normalize $\sigma_Y(f)$, the infinite divergences are no longer guaranteed to cancel.}

In our study, this proved a very real concern.
Overlap reduction functions are given by linear combination of spherical Bessel functions $j_n(\nu)$, where $\nu = 2\pi D f/c$ and $D$ is the distance between detectors~\cite{Flanagan:1993ix,Seto:2007tn}.
The roots of these Bessel functions determine the roots of our signal model in the conventions of Eq.~\eqref{eq:cross-corr-statistic-again} (left side of Fig.~\ref{fig:signal_conventions}), and, correspondingly, the locations of the  divergences in the conventions of Eq.~\eqref{eq:cross-corr-statistic-alt} (right side of Fig.~\ref{fig:signal_conventions}).
For a variety of reasons, different calculations of overlap reduction functions may place these spherical Bessel function roots in fractionally different locations.
If, for example, different users or codes adopt slightly different estimates of detector locations, they will calculate Bessel function roots at slightly different frequencies, and in turn, obtain divergences at slightly different locations when renormalizing their signal model to follow the conventions of Eq.~\eqref{eq:cross-corr-statistic-alt}.
Even more subtly, different code packages will predict Bessel function roots at slightly different locations.
In either case, the signal model $\langle Y(f)\rangle$ produced by a user will contain divergences that do not perfectly match the  divergences in $\sigma^Y(f)$, resulting in a small number of frequencies that have a spurious, outsized impact on SNR and/or likelihood calculations.

In our case, we identified inconsistencies arising from the fact that we compute overlap reduction functions in \texttt{python} using the \texttt{scipy.special.spherical\_jn} function, while the overlap reduction functions computed and pre-applied to $\sigma_Y(f)$ by the LIGO-Virgo-KAGRA collaborations are computed in \texttt{matlab} using its \texttt{besselj} routine.
The inconsistency is illustrated in Fig.~\ref{fig:orf-cancellation-bad}.
The left panel of this figure shows, in blue, the publicly-available O3 Hanford-Virgo uncertainty spectrum $\sigma_Y(f)$ released by the LIGO-Virgo-KAGRA collaboration and, in red, the inverse overlap reduction function $1/\gamma^I(f)$ we compute in \texttt{python}.
Both should exhibit divergences in the same location, but they do not.
The ratio between these curves, the product $\gamma^I(f) \sigma_Y(f) \equiv \sigma(f)$, is what appears in SNR and likelihood calculations; this ratio is shown on the right panel of Fig.~\ref{fig:orf-cancellation-bad}.
We see that the imperfect agreement between overlap reduction functions computed in \texttt{python} versus \texttt{matlab} introduces spurious combs in the ratio, combs that propagate into estimates of the stochastic SNR and likelihood.

In order to avoid such spurious features, when dividing by $\gamma_I(f)$ to match the LIGO-Virgo-KAGRA convention~\eqref{eq:cross-corr-statistic-alt}, it was necessary to install and run the \texttt{matlab}-based \texttt{stochastic.m} package to generate and use the \textit{exact} overlap reduction function that was pre-applied to $\sigma_Y(f)$.
Figure~\ref{fig:orf-cancellation-good} illustrates the successful cancellation between divergences in this case, including the successful recovery of a smooth $\sigma(f) = \gamma^I(f) \sigma_Y(f)$ with no spurious combs.
We note that the disagreement between \texttt{python} and \texttt{matlab} calculations was not simply a matter of different algorithms being used to compute $\gamma^I(f)$; differences persisted even when translating the \texttt{stochastic.m} calculation line-by-line into \texttt{python}.

The different behaviors between Figs.~\ref{fig:orf-cancellation-bad} and \ref{fig:orf-cancellation-good} can have very real effects on stochastic SNR calculations.
Figure~\ref{fig:cumulative_snrs_orf_comparison} again shows the cumulative SNR measured in the Hanford-Virgo baseline during O3 (previously plotted as Fig.~\ref{fig:cumulative-snrs}).
Here, though, we show the cumulative SNRs in two cases: (\textit{i}) using the \texttt{matlab} overlap reduction function to properly cancel the infinite divergences in $\sigma_Y(f)$, and (\textit{ii}) when we use an overlap reduction function computed in \texttt{python} that does not exactly cancel divergences in $\sigma_Y(f)$.
In the second case, the spurious comb that arises in the product $\gamma^I(f)\sigma_Y(f)$ yields discontinuous and significant random jumps in the cumulative SNR.
Somewhat fortuitously, these unphysical jumps nearly cancel one another, returning a total SNR that is comparable in both cases.
This cancelation is purely random, however, and generically it appears that we can expect significant differences in SNR between the two cases, differences that could significantly alter upper limits or even lead to false detection claims.

We emphasize that these issues arise purely from the choice to follow the conventions in Eq.~\eqref{eq:cross-corr-statistic-alt} and adopt uncertainty spectra $\sigma_Y(f) = \sigma(f)/\gamma^I(f)$, a choice that forces downstream analysts to ``undo'' a division by zero.
Accordingly, we recommend the following:
    \begin{itemize}
    \item Ideally, future data products released by the LIGO-Virgo-KAGRA collaboration should follow the conventions of Eq.~\eqref{eq:cross-corr-statistic-again}.
    On the one hand, this convention is mildly more difficult for casual readers to interpret; in the case of an unpolarized stochastic background, the cross-correlation spectrum $\hat C(f)$ is no longer a direct estimator of $\Omega_I(f)$ but instead of the product $\gamma^I(f)\Omega_I(f)$.
    On the other hand, it is no longer necessary for analysts to ``undo'' divisions by zero, an operation that runs the risk of numerical instabilities and spurious detection claims.
    \item Alternatively, public data products should contain all of $\sigma_Y(f)$, $\sigma(f)$, and $\gamma^I(f)$, rather than $\sigma_Y(f)$ alone.
    This allows downstream users to normalize their signal models using exactly the same overlap reduction function used to normalize $\sigma_Y(f)$, mitigating the risk of imperfect cancellation and unphysical results.
    \end{itemize}

\bibliography{master}

 \end{document}